\documentclass[12pt]{scrartcl}
\usepackage{hyperref}
\usepackage{epsfig}
\usepackage[latin1]{inputenc}
\usepackage[T1]{fontenc}
\usepackage{textcomp}
\usepackage[numbers,sort]{natbib}
\usepackage{amsmath}
\usepackage{amsbsy}
\usepackage{amsfonts}
\usepackage{mathptmx}
\usepackage[scaled=0.92]{helvet}
\usepackage{courier}
\usepackage{geometry}
\usepackage{url}
\usepackage[bottom,perpage,flushmargin]{footmisc}

\providecommand{\email}[1]{email: \href{mailto:#1}{\texttt{#1}}}
\providecommand{\dprod}{ \cdot }%
\providecommand{\wprod}{ \wedge }
\providecommand{\pre}[1]{{ }^#1\!}
\renewcommand{\mathrm}[1]{\text{#1}}
\numberwithin{equation}{section}

\title{\vspace{4cm}Can physics laws be derived from monogenic functions?}

\author{José B. Almeida\\ Universidade do Minho, Physics
Department\\
Braga, Portugal, \email{josebda@gmail.com}}

\date{\vspace{-0.5cm}}

\begin{document}

\maketitle



\begin{abstract}
This is a paper about geometry and how one can derive several
fundamental laws of physics from a simple postulate of geometrical
nature. The method uses monogenic functions analysed in the algebra
of 5-dimensional spacetime, exploring the 4-dimensional waves that
they generate. With this method one is able to arrive at equations
of relativistic dynamics, quantum mechanics and electromagnetism.
Fields as disparate as cosmology and particle physics will be
influenced by this approach in a way that the paper only suggests.
The paper provides an introduction to a formalism which shows
prospects of one day leading to a theory of everything and suggests
several areas of future development.
\end{abstract}

\section{Introduction}
The editor's invitation to write a chapter for this book about ether
and the Universe led me to think how my recent work had anything to
do with ether, because the word was never used previously in my
writings. It will become clear in the following sections that the
concept of a privileged frame or absolute motion underlies all the
argument. When one accepts the existence of a preferred frame, the
question of attaching that frame to some observable feature of the
Universe is immediate. This question is addressed in Sec.\
\ref{hyperspherical} but we can anticipate that galaxy clusters are
fixed and can be seen as the anchors for the preferred frame. This
statement seems inconsistent with the observation that clusters of
galaxies move relative to each other but it is resolved invoking an
hyperspherical symmetry in the Universe that is revealed by the
choice of appropriate coordinates.

The relationship between geometry and physics is probably stronger
in the General Theory of Relativity (GTR) than in any other physics
field. It is the author's belief that a perfect theory will
eventually be formulated, where geometry and physics become
indistinguishable, so that the complete understanding of space
properties, together with proper assignments between geometric and
physical entities, will provide all necessary predictions, not only
in relativistic dynamics but in physics as a whole.

We don't have such perfect theory yet, however the author intends to
show that GTR and Quantum Mechanics (QM) can be seen as originating
from monogenic functions in the algebra of the 5-dimensional
spacetime ${G}_{4,1}$. These functions can generate a null
displacement condition, thus reducing the dimensionality by one to
the number of dimensions we are all used to. Besides generating GTR
and QM, the same space generates also 4-dimensional Euclidean space
where dynamics can be formulated and is quite often equivalent to
the relativistic counterpart; Euclidean relativistic dynamics
resembles Fermat's principle extended to 4 dimensions and is thus
designated as 4-Dimensional Optics (4DO).

Our goal is to show how the important equations of physics, such as
relativity equations and equations of quantum mechanics, can be put
under the umbrella of a common mathematical
approach\cite{Almeida05:4, Almeida06:2}. We use geometric algebra as
the framework but introduce monogenic functions with their null
derivatives in order to advance the concept.  Furthermore, we
clarify some previous work in this direction and identify the steps
to take in order to complete this ambitious project.

Since A. Einstein formulated dynamics in 4-dimensional spacetime,
this space is recognized by the vast majority of physicists as being
the best for formulating the laws of physics. However, mathematical
considerations lead to several alternative 4D spaces. For example,
the Euclidean 4-dimensional space of 4DO is equivalent to the 4D
spacetime of GTR when the metric is static, and therefore the
geodesics of one space can be mapped one-to-one with those of the
other. Then one can choose to work in the space that is more
suitable. We build upon previous work by ourselves and by other
authors about null geodesics, regarding the condition that all
material particles must follow null geodesics of 5D space:

\begin{quotation}
{The implication of this for particles is clear: they should travel
on null 5D geodesics. This idea has recently been taken up in the
literature, and has a considerable future. It means that what we
perceive as massive particles in 4D are akin to photons in
5D.}\cite{Wesson05:1}
\end{quotation}

\begin{quotation}
{Accordingly, particles moving on null paths in 5D $(\text{d}S^2 =
0)$ will appear as massive particles moving on timelike paths in 4D
$(\text{d}s^2
> 0)$ \ldots}\cite{Liko03}
\end{quotation}

We actually improve on these null displacement ideas by introducing
the more fundamental monogenic condition, deriving the former from
the latter and establishing a common first principle.

The only postulates in this paper are of a geometrical nature and
can be summarized in the definition of the space we are going to
work with; this is the 4-dimensional null subspace of the
5-dimensional space with signature $(-++++)$. The choice of this
geometric space does not imply any assumption for physical space up
to the point where geometric entities like coordinates and geodesics
start being assigned to physical quantities like distances and
trajectories. Some of those assignments will be made very soon in
the exposition and will be kept consistently until the end in order
to allow the reader some assessment of the proposed geometric model
as a tool for the prediction of physical phenomena. Mapping between
geometry and physics is facilitated if one chooses to work always
with non-dimensional quantities; this is done with a suitable choice
for standards of the fundamental units. From this point onwards all
problems of dimensional homogeneity are avoided through the use of
normalizing factors listed below for all units, defined with
recourse to the fundamental constants: $\hbar \rightarrow$ Planck
constant divided by $2 \pi$, $G \rightarrow$ gravitational constant,
$c \rightarrow$ speed of light and $e \rightarrow$ proton charge.

\begin{center}
\begin{tabular}{c|c|c|c}
Length & Time & Mass & Charge \\
\hline & & & \\

$\displaystyle \sqrt{\frac{G \hbar}{c^3}} $ & $\displaystyle
\sqrt{\frac{G \hbar}{c^5}} $  & $\displaystyle \sqrt{\frac{ \hbar c
}{G}} $  & $e$
\end{tabular}
\end{center}

This normalization defines a system of \emph{non-dimensional units}
(Planck units) with important consequences, namely: 1) All the
fundamental constants, $\hbar$, $G$, $c$, $e$, become unity; 2) a
particle's Compton frequency, defined by $\nu = mc^2/\hbar$, becomes
equal to the particle's mass; 3) the frequent term ${GM}/({c^2 r})$
is simplified to ${M}/{r}$.

5-dimensional space can have amazing structure, providing countless
parallels to the physical world; this paper is just a limited
introductory look at such structure and parallels. The exposition
makes full use of an extraordinary and little known mathematical
tool called geometric algebra (GA), a.k.a.\ Clifford algebra, which
received an important thrust with the works of David Hestenes
\cite{Hestenes84}. A good introduction to GA can be found in
\citet{Gull93} and the following paragraphs use basically the
notation and conventions therein. A complete course on physical
applications of GA can be downloaded from the internet
\cite{Lasenby99}; the same authors published a more comprehensive
version in book form \cite{Doran03}. An accessible presentation of
mechanics in GA formalism is provided by \citet{Hestenes03}. This is
the subject of first section, where some essential GA concepts and
notation are introduced.

Section two deals with monogenic function in flat 5D spacetime,
deriving special relativity and the free particle Dirac equation
from this simple concept. 4DO appears here as a perfect equivalent
to special relativity, where trajectories can be understood as
normals to 4-dimensional plane-like waves. The following section
improves on this by allowing for curved space, introducing the
notion of refractive index tensor. Section five examines the
variational principle applied in both 4DO and GTR spaces to justify
the equivalence of geodesics between the two spaces for static
metrics. Refractive index is then related to its sources and the
sources tensor is defined. The case of a central mass is examined
and the links to Schwarzschild's metric are thoroughly discussed.
Electromagnetism and electrodynamics are formulated as particular
cases of refractive index in section seven and the sources tensor is
here related to a current vector. The next section introduces the
hypothesis of an hyperspherical symmetry in the Universe, which
would call for the use of hyperspherical coordinates; the
consequences for cosmology would include a complete dismissal of
dark matter for a flat rate Hubble expansion. Before the conclusion,
section nine shows how the monogenic condition is effective in
generating an $SU(4)$ symmetry group and makes some advances towards
a relation with the standard model of particle physics.

\section{Introduction to geometric algebra \label{somealg}}
Geometric algebra is not usually taught in university courses and
its presence in the literature is scarce; good reference works are
\cite{Doran03, Hestenes84, Lasenby99}. We will concentrate on the
algebra of 5-dimensional spacetime because this will be our main
working space; this algebra incorporates as subalgebras those of the
usual 3-dimensional Euclidean space, Euclidean 4-space and Minkowski
spacetime. We begin with the simpler 5D flat space and progress to a
5D spacetime of general curvature.

The geometric algebra ${G}_{4,1}$ of the hyperbolic 5-dimensional
space we consider is generated by the coordinate frame of
orthonormal basis vectors $\sigma_\alpha $ such that
\begin{eqnarray}
\label{eq:basis}
    && (\sigma_0)^2  = -1, \nonumber \\
    && (\sigma_i)^2 =1, \\
    && \sigma_\alpha \dprod \sigma_\beta  =0, \quad \alpha \neq \beta. \nonumber
    \nonumber
\end{eqnarray}
Note that the English characters i, j, k range from 1 to 4 while the
Greek characters $\alpha, \beta, \gamma$ range from 0 to 4. See the
Appendix \ref{indices} for the complete notation convention used.

Any two basis vectors can be multiplied, producing the new entity
called a bivector. This bivector is the geometric product or, quite
simply, the product; this product is distributive. Similarly to the
product of two basis vectors, the product of three different basis
vectors produces a trivector and so forth up to the fivevector,
because five is the dimension of space.

We will simplify the notation for basis vector products using
multiple indices, i.e.\ $\sigma_\alpha \sigma_\beta \equiv
\sigma_{\alpha\beta}.$ The algebra is 32-dimensional and is spanned
by the basis
\begin{itemize}
\item 1 scalar, { $1$},
\item 5 vectors, { $\sigma_\alpha$},
\item 10 bivectors (area), { $\sigma_{\alpha\beta}$},
\item 10 trivectors (volume), { $\sigma_{\alpha\beta\gamma}$},
\item 5 tetravectors (4-volume), { $\mathrm{i} \sigma_\alpha $},
\item 1 pseudoscalar (5-volume), { $\mathrm{i} \equiv
\sigma_{01234}$}.
\end{itemize}
Several elements of this basis square to unity:
\begin{equation}
\label{eq:positive}
    (\sigma_i)^2 =  (\sigma_{0i})^2=
    (\sigma_{0i j})^2 =(\mathrm{i}\sigma_0)^2 =1.
\end{equation}
It is easy to verify the equations above; suppose we want to check
that $(\sigma_{0i j})^2 = 1$. Start by expanding the square and
remove the compact notation $(\sigma_{0i j})^2 = \sigma_0 \sigma_i
\sigma_j \sigma_0 \sigma_i \sigma_j$, then swap the last $\sigma_j$
twice to bring it next to its homonymous; each swap changes the
sign, so an even number of swaps preserves the sign: $(\sigma_{0i
j})^2 = \sigma_0 \sigma_i (\sigma_j)^2 \sigma_0 \sigma_i$. From the
third equation (\ref{eq:basis}) we know that the squared vector is
unity and we get successively $(\sigma_{0i j})^2 = \sigma_0 \sigma_i
\sigma_0 \sigma_i = - (\sigma_0)^2 (\sigma_i)^2 = - (\sigma_0)^2 $;
using the first equation (\ref{eq:basis}) we get finally
$(\sigma_{0i j})^2 = 1$ as desired.

The remaining basis elements square to $-1$ as can be verified in a
similar manner:
\begin{equation}
    (\sigma_0)^2 = (\sigma_{ij})^2 = (\sigma_{ijk})^2 =
    (\mathrm{i}\sigma_i)^2 = \mathrm{i}^2=-1.
\end{equation}
Note that the pseudoscalar $\mathrm{i}$ commutes with all the other
basis elements while being a square root of $-1$; this makes it a
very special element which can play the role of the scalar imaginary
in complex algebra.

We can now address the geometric product of any two vectors $a =
a^\alpha \sigma_\alpha$ and $b = b^\beta \sigma_\beta$ making use of
the distributive property
\begin{equation}
    ab = \left(-a^0 b^0 + \sum_i a^i b^i \right) + \sum_{\alpha \neq \beta}
    a^\alpha b^\beta \sigma_{\alpha \beta};
\end{equation}
and we notice it can be decomposed into a symmetric part, a scalar
called the inner or interior product, and an anti-symmetric part, a
bivector called the outer or exterior product.
\begin{equation}
    ab = a \dprod b + a \wprod b,~~~~ ba = a \dprod b - a \wprod b.
\end{equation}
Reversing the definition one can write inner and outer products as
\begin{equation}
    a \dprod b = \frac{1}{2}\, (ab + ba),~~~~ a \wprod b = \frac{1}{2}\, (ab -
    ba).
\end{equation}
The inner product is the same as the usual ''dot product,'' the only
difference being in the negative sign of the $a_0 b_0$ term; this is
to be expected and is similar to what one finds in special
relativity. The outer product represents an oriented area; in
Euclidean 3-space it can be linked to the "cross product" by the
relation $\mathrm{cross}(\mathbf{a},\mathbf{b}) = - \sigma_{123}
\mathbf{a} \wprod \mathbf{b}$; here we introduced bold characters
for 3-dimensional vectors and avoided defining a symbol for the
cross product because we will not use it again. We also used the
convention that interior and exterior products take precedence over
geometric product in an expression.

When a vector is operated with a multivector the inner product
reduces the grade of each element by one unit and the outer product
increases the grade by one. We will generalize the definition of
inner and outer products below; under this generalized definition
the inner product between a vector and a scalar produces a vector.
Given a multivector $a$ we refer to its grade-$r$ part by writing
$<\!a\!>_r$; the scalar or grade zero part is simply designated as
$<\!a\!>$. By operating a vector with itself we obtain a scalar
equal to the square of the vector's length
\begin{equation}
    a^2 = aa = a \dprod a + a \wprod a = a \dprod a.
\end{equation}
The definitions of inner and outer products can be extended to
general multivectors
\begin{eqnarray}
    a \dprod b &=& \sum_{\alpha,\beta} \left<<\!a\!>_\alpha \;
    <\!b\!>_\beta \right>_{|\alpha-\beta|},\\
    a \wprod b &=& \sum_{\alpha,\beta} \left<<\!a\!>_\alpha \;
    <\!b\!>_\beta \right>_{\alpha+\beta}.
\end{eqnarray}
Two other useful products are the scalar product, denoted as
$<\!ab\!>$ and commutator product, defined by
\begin{equation}
    a \times b = ab - ba.
\end{equation}
In mixed product expressions we will always use the convention that
inner and outer products take precedence over geometric products,
thus reducing the number of parenthesis.

We will encounter exponentials with multivector exponents; two
particular cases of exponentiation are specially important. If $u$
is such that $u^2 = -1$ and $\theta$ is a scalar
\begin{eqnarray}
   \mathrm{e}^{u \theta} &=& 1 + u \theta -\frac{\theta^2}{2!} - u
    \frac{\theta^3}{3!} + \frac{\theta^4}{4!} + \ldots  \nonumber \\
    &=& 1 - \frac{\theta^2}{2!} +\frac{\theta^4}{4!}- \ldots \{=
    \cos \theta \} + \nonumber \\
    && + u \theta - u \frac{\theta^3}{3!} + \ldots \{= u \sin
    \theta\} \\
    &=&  \cos \theta + u \sin \theta. \nonumber
\end{eqnarray}
Conversely if $h$ is such that $h^2 =1$
\begin{eqnarray}
    \mathrm{e}^{h \theta} &=& 1 + h \theta +\frac{\theta^2}{2!} + h
    \frac{\theta^3}{3!} + \frac{\theta^4}{4!} + \ldots  \nonumber \\
    &=& 1 + \frac{\theta^2}{2!} +\frac{\theta^4}{4!}+ \ldots \{=
    \cosh \theta \} + \nonumber \\
    && + h \theta + h \frac{\theta^3}{3!} + \ldots \{= h \sinh
    \theta\}  \\
    &=&  \cosh \theta + h \sinh \theta. \nonumber
\end{eqnarray}
The exponential of bivectors is useful for defining rotations; a
rotation of vector $a$ by angle $\theta$ on the $\sigma_{12}$ plane
is performed by
\begin{equation}
    a' = \mathrm{e}^{\sigma_{21} \theta/2} a
    \mathrm{e}^{\sigma_{12} \theta/2}= \tilde{R} a R;
\end{equation}
the tilde denotes reversion and reverses the order of all products.
As a check we make $a = \sigma_1$
\begin{eqnarray}
    \mathrm{e}^{-\sigma_{12} \theta/2} \sigma_1
    \mathrm{e}^{\sigma_{12} \theta/2} &=&
    \left(\cos \frac{\theta}{2} - \sigma_{12}
    \sin \frac{\theta}{2}\right) \sigma_1 \ast \nonumber \\
    &&\ast \left(\cos \frac{\theta}{2} + \sigma_{12} \sin
    \frac{\theta}{2}\right)  \\
    &=& \cos \theta \sigma_1 + \sin \theta \sigma_2. \nonumber
\end{eqnarray}
Similarly, if we had made $a = \sigma_2,$ the result would have been
$-\sin \theta \sigma_1 + \cos \theta \sigma_2.$

If we use $B$ to represent a bivector whose plane is normal to
$\sigma_0$ and define its norm by $|B| = (B \tilde{B})^{1/2},$ a
general rotation in 4-space is represented by the rotor
\begin{equation}
    R \equiv e^{-B/2} = \cos\left(\frac{|B|}{2}\right) -  \frac{B}{|B|}
    \sin\left(\frac{|B|}{2}\right).
\end{equation}
The rotation angle is $|B|$ and the rotation plane is defined by
$B.$ A rotor is defined as a unitary even multivector (a multivector
with even grade components only) which squares to unity; we are
particularly interested in rotors with bivector components. It is
more general to define a rotation by a plane (bivector) then by an
axis (vector) because the latter only works in 3D while the former
is applicable in any dimension. When the plane of bivector $B$
contains $\sigma_0$, a similar operation does not produce a simple
rotation but produces a boost, eventually combined with a rotation.
Take for instance $B = \sigma_{01} \theta/2$ and define the
transformation operator $T = \exp( B)$; a transformation of the
basis vector $\sigma_0$ produces
\begin{eqnarray}
    a' &=& \tilde{T} \sigma_0 T = \mathrm{e}^{-\sigma_{01}\theta/2}
    \sigma_0 \mathrm{e}^{\sigma_{01}\theta/2} \nonumber \\
    &=& \left(\cosh \frac{\theta}{2} - \sigma_{01}
    \sinh \frac{\theta}{2}\right) \sigma_0 \ast \nonumber \\
    &&\ast\left(\cosh \frac{\theta}{2} + \sigma_{01} \sinh
    \frac{\theta}{2}\right) \\
    &=& \cosh \theta \sigma_0 + \sinh \theta \sigma_1. \nonumber
\end{eqnarray}

In 5-dimensional spacetime of general curvature, we introduce 5
coordinate frame vectors $g_\alpha$, the indices follow the
conventions set forth in Appendix \ref{indices}. We will also assume
this spacetime to be a metric space whose metric tensor is given by
\begin{equation}
\label{eq:metrictens}
    g_{\alpha \beta} = g_\alpha \dprod g_\beta;
\end{equation}
the double index is used with $g$ to denote the inner product of
frame vectors and not their geometric product. The space signature
is $(-++++)$, which amounts to saying that $g_{00} < 0$ and $g_{ii}
>0$. A reciprocal frame is defined by the condition
\begin{equation}
   \label{eq:recframe}
    g^\alpha \dprod g_\beta = {\delta^\alpha}_\beta.
\end{equation}
Defining $g^{\alpha \beta}$ as the inverse of $g_{\alpha \beta}$,
the matrix product of the two must be the identity matrix; using
Einstein's summation convention this is
\begin{equation}
    g^{\alpha \gamma} g_{\gamma \beta} = {\delta^\alpha}_\beta.
\end{equation}
Using the definition (\ref{eq:metrictens}) we have
\begin{equation}
    \left(g^{\alpha \gamma} g_\gamma \right)\dprod g_\beta =
    {\delta^\alpha}_\beta;
\end{equation}
comparing with Eq.\ (\ref{eq:recframe}) we  determine $g^\alpha$with
\begin{equation}
    g^\alpha = g^{\alpha \gamma} g_\gamma.
\end{equation}

If the coordinate frame vectors can be expressed as a linear
combination of the orthonormed ones, we have
\begin{equation}
    \label{eq:indexframemain}
    g_\alpha = {n^\beta}_\alpha \sigma_\beta,
\end{equation}
where ${n^\beta}_\alpha$ is called the \emph{refractive index
tensor} or simply the \emph{refractive index}; its 25 elements can
vary from point to point as a function of the
coordinates.\cite{Almeida04:4, Almeida06:2} When the refractive
index is the identity, we have $g_\alpha = \sigma_\alpha$ for the
main or direct frame and $g^0 = -\sigma_0$, $g^i = \sigma_i$ for the
reciprocal frame, so that Eq.\ (\ref{eq:recframe}) is verified. In
this work we will not consider spaces of general curvature but only
those satisfying condition (\ref{eq:indexframemain}).

The first use we will make of the reciprocal frame is for the
definition of two derivative operators. In flat space we define the
vector derivative
\begin{equation}
\label{eq:nabla}
    \nabla = \sigma^\alpha\partial_\alpha.
\end{equation}
It will be convenient, sometimes, to use vector derivatives in
subspaces of 5D space; these will be denoted by an upper index
before the $\nabla$ and the particular index used determines the
subspace to which the derivative applies; For instance
$\pre{m}\nabla = \sigma^m \partial_m = \sigma^1 \partial_1 +
\sigma^2 \partial_2 + \sigma^3 \partial_3.$ In 5-dimensional space
it will be useful to split the vector derivative into its time and
4-dimensional parts
\begin{equation}
    \nabla = -\sigma_0\partial_t + \sigma^i \partial_i
    = -\sigma_0\partial_t
    + \pre{i}\nabla.
\end{equation}

The second derivative operator is the covariant derivative,
sometimes called the \emph{Dirac operator}, and it is defined in the
reciprocal frame $g^\alpha$
\begin{equation}
\label{eq:covariant}
    \mathrm{D} = g^\alpha\partial_\alpha.
\end{equation}
Taking into account the definition of the reciprocal frame
(\ref{eq:recframe}), we see that the covariant derivative is also a
vector. In cases such as those we consider in this work, where there
is a refractive index, it will be possible to define both
derivatives in the same space.

We define also second order differential operators, designated
Laplacian and covariant Laplacian respectively, resulting from the
inner product of one derivative operator by itself. The square of a
vector is always a scalar and the vector derivative is no exception,
so the Laplacian is a scalar operator, which consequently acts
separately in each component of a multivector. For $4+1$ flat space
it is
\begin{equation}
    \nabla^2 = -\frac{\partial^2}{\partial t^2} + \pre{i}\nabla^2.
\end{equation}
One sees immediately that a 4-dimensional wave equation is obtained
by zeroing the Laplacian of some function
\begin{equation}
    \label{eq:4dwavemain}
    \nabla^2 \psi = \left(-\frac{\partial^2}{\partial t^2} +
    \pre{i}\nabla^2\right)\psi = 0.
\end{equation}
This procedure will be used in the next section for the derivation
of special relativity and will be extended later to general curved
spaces.

\section{Monogenic functions and waves in flat space \label{dynamics}}

It turns out that there is a class of functions of great importance,
called \emph{monogenic functions},\cite{Doran03} characterized by
having null vector derivative; a function $\psi$ is monogenic if and
only if
\begin{equation}
    \label{eq:monogenic}
    \nabla \psi = 0.
\end{equation}
A monogenic function has by necessity null Laplacian, as can be seen
by dotting Eq.\ (\ref{eq:monogenic}) with $\nabla$ on the left. We
are then allowed to write
\begin{equation}
    \label{eq:solution}
    \sum_i \partial_{ii} \psi = \partial_{00} \psi.
\end{equation}
This can be recognized as a wave equation in the 4-dimensional space
spanned by $\sigma_i$ which will accept plane wave type solutions of
the general form
\begin{equation}
    \label{eq:psidef}
    \psi = \psi_0 \mathrm{e}^{\mathrm{i} (p_\alpha x^\alpha + \delta)},
\end{equation}
where $\psi_0$ is an amplitude whose characteristics we shall not
discuss for now, $\delta$ is a phase angle and $p_\alpha$ are
constants such that
\begin{equation}
    \label{eq:pnull}
    \sum_i (p_i)^2 - (p_0)^2 = 0.
\end{equation}

By setting the argument of $\psi$ constant in Eq.\ (\ref{eq:psidef})
and differentiating we can get the differential equation
\begin{equation}
    \label{eq:nullcond}
    p_\alpha \mathrm{d}x^\alpha =  0.
\end{equation}
The first member can equivalently be written as the inner product of
the two vectors $p \dprod \mathrm{d}x = 0$, where $p = \sigma^\alpha
p_\alpha$. In 5D hyperbolic space the inner product of two vectors
can be null when the vectors are perpendicular but also when the two
vectors are null; since we have established that $p$ is a null
vector, Eq.\ (\ref{eq:nullcond}) can be satisfied either by
$\mathrm{d}x$ normal to $p$ or by $(\mathrm{d}x)^2 = 0$. In the
former case the condition describes a 3-volume called wavefront and
in the latter case it describes the wave motion. Notice that the
wavefronts are not surfaces but volumes, because we are working with
4-dimensional waves.

The condition describing wave motion can be expanded as
\begin{equation}
    \label{eq:wavemotion}
    -(\mathrm{d}x^0)^2 + \sum (\mathrm{d}x^i)^2 = 0.
\end{equation}
This is a purely scalar equation and can be manipulated as such,
which means we are allowed to rewrite it with any chosen terms in
the second member; some of those manipulations are particularly
significant. Suppose we decide to isolate $(\mathrm{d}x^4)^2$ in the
first member: $(\mathrm{d}x^4)^2 = (\mathrm{d}x^0)^2 - \sum
(\mathrm{d}x^m)^2$. We can then rename coordinate $x^4$ as $\tau$,
to get the interval squared of special relativity for space-like
displacements
\begin{equation}
    \mathrm{d}\tau^2 = (\mathrm{d}x^0)^2 - \sum
(\mathrm{d}x^m)^2.
\end{equation}
We have thus derived the space-like part of special relativity as a
consequence of monogeneity in 5D hyperbolic space and simultaneously
justified the physical interpretation for coordinates $x^0$ and
$x^4$ as time and proper time, respectively.

A different manipulation of Eq.\ (\ref{eq:wavemotion}) has great
significance because it leads to the 4DO concept.\cite{Almeida02:2,
Almeida01} If we isolate $(\mathrm{d}x^0)^2$ and replace $x^0$ by
the letter $t$, we see that time becomes the interval in Euclidean
4D space
\begin{equation}
\label{eq:twospaces}
    \mathrm{d}t^2 = \sum (\mathrm{d}x^i)^2.
\end{equation}
From this we conclude that the monogenic condition produces plane
waves whose wavefronts are 3D volumes but can be represented by
wavefront normals, just as it happens in standard optics with
electromagnetic waves.

Several readers may be worried with the fact that proper time is a
line integral and not a coordinate in special relativity and so
$\mathrm{d} \tau$ should not be allowed to appear on the rhs of the
equation. To this we will argue that the manipulations we have done,
collapsing 5D spacetime into 4 dimensions through a null
displacement condition and then promoting one of the coordinates
into interval, is exactly equivalent to the process of defining a
light cone in Minkowski spacetime and then applying Fermat's
principle to define an Euclidean 3D metric on the light cone; we
have just upgraded the procedure by including one extra dimension.

The Dirac equation can also be derived from the monogenic condition
but since it appears formulated in terms of matrices in all
textbooks we will have to rewrite Eq.\ (\ref{eq:monogenic}) also in
terms of matrices, so that our GA manipulations can also be
understood as matrix operations. This is easily achieved if we
assign our frame vectors to Dirac matrices that square to the the
identity matrix or minus the identity matrix as appropriate; the
following list of assignments can be used but others would be
equally effective\footnote{There are 16 possible $4 \ast 4$ Dirac
matrices,\cite{DiracMat} of which we must choose 5 such that
$(\sigma_0)^2 = -I$, $(\sigma_i)^2 =I$ and $\sigma_\alpha
\sigma_\beta = -\sigma_\beta \sigma_\alpha$, for $\alpha \neq
\beta$; the present choice will simplify our symmetry discussions
further along.}
\begin{equation}
\begin{split}
\label{eq:diracmatrix}
   \sigma^0 &\equiv   \begin{pmatrix} \mathrm{i} & 0 & 0 & 0  \\
    0 & -\mathrm{i} & 0 & 0 \\ 0 & 0 & \mathrm{i} & 0 \\ 0 & 0 & 0 &
    -\mathrm{i}
    \end{pmatrix},~~
    \sigma^1 \equiv  \begin{pmatrix} 0 & 0 & 0 & 1 \\
    0 & 0 & -1 & 0 \\ 0 & -1 & 0 & 0 \\ 1 & 0 & 0 & 0
    \end{pmatrix}, \\
    \sigma^2 &\equiv   \begin{pmatrix} 0 & \mathrm{i} & 0 & 0 \\
    -\mathrm{i} & 0 & 0 & 0 \\ 0 & 0 & 0 & -\mathrm{i} \\ 0 & 0 & \mathrm{i} & 0
    \end{pmatrix},~~
    \sigma^3 \equiv \begin{pmatrix} 0 & 1 & 0 & 0 \\
    1 & 0 & 0 & 0 \\ 0 & 0 & 0 & 1 \\ 0 & 0 & 1 & 0
    \end{pmatrix},  \\
    \sigma^4 &\equiv   \begin{pmatrix} 0 & 0 & 0 & -\mathrm{i} \\
    0 & 0 & \mathrm{i} & 0 \\ 0 & -\mathrm{i} & 0 & 0 \\ \mathrm{i} & 0 & 0 & 0
    \end{pmatrix} .
\end{split}
\end{equation}
There is no need to adopt different notations to refer to the frame
vectors or to their matrix counterparts because the context will
usually be sufficient to determine what is meant.

We can check that matrices $\sigma^\alpha$ form an orthonormal basis
of 5D space by defining the inner product of square matrices as
\begin{equation}
    A \dprod B = \frac{A B + B A}{2}.
\end{equation}
It will then be possible to verify that the inner product of any two
different $\sigma$-matrices is null, $(\sigma^0)^2 = -I$ and
$(\sigma^i)^2 = I$; these are the conditions defining an orthonormal
basis expressed in matrix form. A more formal approach to this
subject would lead us to invoke the isomorphism between the complex
algebra of $4 \ast 4$ matrices and Clifford algebra $G_{4,1}$, the
geometric algebra of 5D spacetime.\cite{Lounesto01}.

It will now be convenient to expand the monogenic condition
(\ref{eq:monogenic}) as $(\sigma^\mu \partial_\mu + \sigma^4
\partial_4) \psi = 0$. If this is applied to the solution
(\ref{eq:psidef}) and the derivative with respect to $x^4$ is
evaluated we get
\begin{equation}
    (\sigma^\mu \partial_\mu + \sigma^4 \mathrm{i} p_4) \psi = 0.
\end{equation}
Let us now multiply both sides of the equation on the left by
$\sigma^4$ and note that matrix $\sigma^4 \sigma^0$ squares to the
identity while the 3 matrices $\sigma^4 \sigma^m$ square to minus
identity; we rename these products as $\gamma$-matrices in the form
$\gamma^\mu = \sigma^4 \sigma^\mu$. Rewriting the equation in this
form we get
\begin{equation}
\label{eq:Diracgamma}
    (\gamma^\mu \partial_\mu + \mathrm{i} p_4) \psi = 0.
\end{equation}
The only thing this equation needs to be recognized as Dirac's is
the replacement of $p_4$ by the particle's mass $m$; simultaneously
we assign the energy $E$ to $p_0$ and $3D$ momentum $\mathbf{p}$ to
$\sigma^m p_m$.

We turn now our attention to the amplitude $\psi_0$ in Eq.\
(\ref{eq:psidef}) because we know that the Dirac equation accepts
solutions which are spinors and we want to find out their
equivalents in our formulation. Applying the monogenic condition to
Eq.\ (\ref{eq:psidef}) we see that the following equation must be
verified
\begin{equation}
    \psi_0 (\sigma^\alpha p_\alpha) = 0.
\end{equation}
If the $\sigma$s are interpreted as matrices, remembering that $p$
is null, the only way the equation can be verified is by $\psi_0$
being some constant multiplied by the matrix in parenthesis, which
is a matrix representation of $p$. We can set the multiplying
constant to unity and $\psi_0$ becomes equal to $p$; the
wavefunction $\psi$ can then be interpreted as a Dirac spinor.  The
{wave} function in Eq.\ (\ref{eq:psidef}) can now be given a
different form, taking in consideration the previous assignments
\begin{equation}
    \label{eq:fermion}
    \psi = A(\sigma_4 m + \mathbf{p} \mp \sigma_0 E)
     \mathrm{e}^{u (\pm E t + \mathbf{p}\cdot \mathbf{x} + m \tau + \delta)};
\end{equation}
where $A$ is the amplitude and $\mathbf{x} = \sigma_m x^m$ is the
3-dimensional position.

In order to separate \emph{left} and \emph{right} spinor components
we use a technique adapted from Ref.\ \cite{Doran03}. We choose an
arbitrary $4 \times 4$ matrix which squares to identity, for
instance $\sigma_4$, with which we form the two idempotent matrices
$(I + \sigma_4)/2$ and $(I - \sigma_4)/2$.\footnote{Matrix
$\sigma^4$ is the same as matrix $\gamma^5 = \mathrm{i}\gamma^0
\gamma^1 \gamma^2 \gamma^3 $.} These matrices are called idempotents
because they reproduce themselves when squared. These idempotents
absorb any $\sigma_4$ factor; as can be easily checked $(I +
\sigma_4) \sigma_4 = (I + \sigma_4)$ and $(I - \sigma_4) \sigma_4 =
- (I - \sigma_4)$.

Obviously we can decompose the wavefunction $\psi$ as
\begin{equation}
    \psi =   \psi \frac{I + \sigma_4}{2} + \psi \frac{I -
    \sigma_4}{2} = \psi_+ + \psi_-.
\end{equation}
This apparently trivial decomposition produces some surprising
results due to the following relations
\begin{eqnarray}
    \mathrm{e}^{\mathrm{i} \theta} (I + \sigma_4) &=&
    (\cos \theta + \mathrm{i} \sin \theta) (I + \sigma_4) \nonumber \\
    &=& (I \cos \theta + \mathrm{i} \sigma_4 \sin \theta)
    (I + \sigma_4) \\
    &=& \mathrm{e}^{\mathrm{i}\sigma_4 \theta} (I + \sigma_4).
    \nonumber
\end{eqnarray}
and similarly
\begin{equation}
\label{eq:leftright}
    \mathrm{e}^{\mathrm{i} \theta} (I - \sigma_4)=
    \mathrm{e}^{-\mathrm{i} \sigma_4 \theta} (I - \sigma_4).
\end{equation}
If we had chosen a different idempotent the result would have been
similar; we will see how the various idempotents are arranged in a
symmetry group and it has been argued that they may be related to
elementary particles.\cite{Almeida05:1}

\section{Relativistic dynamics}
When working in curved spaces the monogenic condition is naturally
modified, replacing the vector derivative $\nabla$ with the
covariant derivative $\mathrm{D}$. A generalized monogenic function
is then a function that verifies the equation
\begin{equation}
\label{eq:genmonogenic}
    \mathrm{D} \psi = 0.
\end{equation}
  Similarly to what
happens in flat space, the covariant Laplacian is a scalar and a
monogenic function must verify the second order differential
equation
\begin{equation}
\label{eq:curvwave}
    \mathrm{D}^2 \psi = 0.
\end{equation}
It is possible to write a general expression for the covariant
Laplacian in terms of the metric tensor components (see
\cite[Section 2.11]{Arfken95}) but we will consider only situations
where that complete general expression is not needed.

When Eq.\ (\ref{eq:genmonogenic}) is multiplied on the left by
$\mathrm{D}$, we are applying second derivatives to the function,
but we are simultaneously applying first order derivatives to the
reciprocal frame vectors present in the definition of $\mathrm{D}$
itself. We can simplify the calculations if the variations of the
frame vectors are taken to be much slower than those of function
$\psi$ so that frame vector derivatives can be neglected. With this
approximation, the covariant Laplacian becomes $\mathrm{D}^2 =
g^{\alpha \beta}
\partial_{\alpha \beta}$ and Eq.\ (\ref{eq:curvwave}) can be written
\begin{equation}
\label{eq:curvwave2}
   g^{\alpha \beta} \partial_{\alpha \beta} \psi = 0.
\end{equation}
This equation can have a solution of the type given by Eq.\
(\ref{eq:psidef}) if again the derivatives of $p_\alpha$ are
neglected. This approximation is usually of the same order as the
former one and should not be seen as a second restriction. Inserting
Eq.\ (\ref{eq:psidef}) one sees that it is a solution if
\begin{equation}
    g^{\alpha \beta} p_\alpha p_\beta =0.
\end{equation}
This equation is the curved space equivalent to Eq.\
(\ref{eq:pnull}) and it means that the square of vector $p =
g^\alpha p_\alpha$ is zero, that is, $p$ is a vector of zero length;
for this reason it is called a null vector or \emph{nilpotent}.
Vector $p$ is the \emph{momentum vector} and should not be confused
with 4-dimensional conjugate momentum vectors defined below.

We arrive again at Eq.\ (\ref{eq:nullcond}) and the condition
describing 4D wave motion can be expanded as
\begin{equation}
    \label{eq:wavemotion2}
    g_{\alpha \beta} \mathrm{d}x^\alpha
    \mathrm{d}x^\beta = 0.
\end{equation}
This condition effectively reduces the spatial dimension to four but
the resulting space is non-metric because all displacements have
zero length. We will remove this difficulty by considering two
special cases. First let us assume that vector $g_0$ is normal to
the other frame vectors so that all $g_{0i}$ factors are zeroed;
condition (\ref{eq:wavemotion2}) becomes
\begin{equation}
\label{eq:nullspecial}
    g_{00}(\mathrm{d}x^0)^2 +
    g_{ij}\mathrm{d}x^i \mathrm{d}x^j = 0.
\end{equation}
All the terms in this equation are scalars and we are allowed to
rewrite it with $(\mathrm{d}x^0)^2$ in the lhs
\begin{equation}
\label{eq:4dometric}
    (\mathrm{d}x^0)^2 = -\frac{g_{ij}}{g_{00}}\, \mathrm{d}x^i
    \mathrm{d}x^j.
\end{equation}
We could have arrived at the same result by defining a 4-dimensional
displacement vector
\begin{equation}
\label{eq:velocity}
    \mathrm{d}x^0 v = \frac{-1}{\sqrt{g_{00}}}\, g_i \mathrm{d}x^i;
\end{equation}
and then squaring it to evaluate its length; $v$ is a unit vector
called velocity because its definition is similar to the usual
definition of 3-dimensional velocity; its components are
\begin{equation}
    v_i =  \frac{\mathrm{d}x^i}{\mathrm{d}x^0}.
\end{equation}
Being unitary, the velocity can be obtained by a rotation of the
$\sigma_4$ frame vector
\begin{equation}
    v = \tilde{R}\sigma_4 R.
\end{equation}
The rotation angle is a measure of the 3-dimensional velocity
component. A null angle corresponds to $v$ directed along $\sigma_4$
and null 3D component, while a $\pi/2$ angle corresponds to the
maximum possible 3D component. The idea that physical velocity can
be seen as the 3D component of a unitary 4D vector has been explored
in several papers but see \cite{Almeida01:4}.

Equation (\ref{eq:velocity}) projects the original 5-dimensional
space into an Euclidean signature 4 dimensional space, where an
elementary displacement is given by the variation of coordinate
$x^0$. In the particular case where $g_0 = \sigma_0$ the
displacement vector simplifies to $\mathrm{d}x^0 v = g_i
\mathrm{d}x^i$ and we can see clearly that the signature is
Euclidean because the four $g_i$ have positive norm. Although it has
not been mentioned, we have assumed that none of the frame vectors
is a function of coordinate $x^0$.

Returning to Eq.\ (\ref{eq:nullspecial}) we can now impose the
condition that $g_4$ is normal to the other frame vectors in order
to isolate $(\mathrm{d}x^4)^2$ instead of $(\mathrm{d}x^0)^2$, as we
did before;
\begin{equation}
\label{eq:grmetric}
    (\mathrm{d}x^4)^2 = - \frac{g_{\mu\nu}}{g_{44}}\,
    \mathrm{d}x^\mu
    \mathrm{d}x^\nu .
\end{equation}
We have now projected onto 4-dimensional space with signature
$(+---)$, known as Minkowski signature. In order to check this
consider again the special case with $g_0 = \sigma_0$ and the
equation becomes
\begin{equation}
    (\mathrm{d}x^4)^2 = \frac{1}{g_{44}}\, (\mathrm{d}x^0)^2
    - \frac{g_{mn}}{g_{44}}\, \mathrm{d}x^m
    \mathrm{d}x^n ;
\end{equation}
the diagonal elements $g_{ii}$ are necessarily positive, which
allows a verification of Minkowski signature. Contrary to what
happened in the previous case, we cannot now obtain
$(\mathrm{d}x^4)^2$ by squaring a vector but we can do it by
consideration of the bivector
\begin{equation}
\label{eq:dx4}
    \mathrm{d}x^4 \nu  = \frac{1}{\sqrt{g_{44} g^{44}}}\, g_{\mu}g^4 \mathrm{d}x^\mu.
\end{equation}
All the products $g_{\mu}g^4$ are bivectors because we imposed $g_4$
to be normal to the other frame vectors. When $(\mathrm{d}x^4)^2$ is
evaluated by an inner product we notice that $g_0 g^4$ has positive
square while the three $g_m g^4$ have negative square, ensuring that
a Minkowski signature is obtained. Naturally we have to impose the
condition that none of the frame vectors depends on $x^4$. Bivector
$\nu$ is such that $\nu^2 = \nu \nu =1$ and it  can be obtained by a
Lorentz transformation of bivector $\sigma_{04}$.
\begin{equation}
    \nu = \tilde{T} \sigma_{04} T,
\end{equation}
where $T$ is of the form $T = \exp(B)$ and $B$ is a bivector whose
plane is normal to $\sigma_4$. Note that $T$ is a pure rotation when
the bivector plane is normal to both $\sigma_0$ and $\sigma_4$.

In special relativity it is usual to work in a space spanned by an
orthonormed frame of vectors $\gamma_\mu$ such that $(\gamma_0)^2 =
1$ and $(\gamma_m)^2 = -1$, producing the desired Minkowski
signature \cite{Doran03}. The geometric algebra of this space is
isomorphic to the even sub-algebra of $G_{4,1}$ and so the area
element $\mathrm{d}x^4 \nu$ (\ref{eq:dx4}) can be reformulated as a
vector called relativistic 4-velocity. The four $\gamma$ bivectors
are defined in a similar way to the $\gamma$ matrices used in Eq.\
(\ref{eq:Diracgamma}), which is to be expected from the isomorphism
between geometric and matrix algebras already mentioned.

Equations (\ref{eq:4dometric}) and (\ref{eq:grmetric}) define two
alternative 4-dimensional spaces, those of \emph{4-dimensional
optics (4DO)}, with metric tensor $-g_{ij}/g_{00}$ and \emph{general
theory of relativity (GTR)} with metric tensor $-g_{\mu
\nu}/g_{44}$, respectively; in the former $x^0$ is an affine
parameter while in the latter it is $x^4$ that takes such role. In
fact Eq.\ (\ref{eq:grmetric}) only covers the spacelike part of GTR
space, because $(\mathrm{d}x^4)^2$ is necessarily non-negative.
Naturally there is the limitation that the frame vectors are
independent of both $x^0$ and $x^4$, equivalent to imposing a static
metric, and also that $g_{0i} = g_{\mu 4} =0$. Provided the metric
is static, the geodesics of 4DO can be mapped one-to-one with
spacelike geodesics of GTR and we can choose to work on the space
that best suits us for free fall dynamics. For a physical
interpretation of geometric relations it will frequently be
convenient to assign new designations to the 5D coordinates that
acquire the role of affine parameter in the null subspace. We recall
the assignments $x^0 \equiv t$ and $x^4 \equiv \tau$; total
derivatives with respect to these coordinates will receive a special
notation: $\mathrm{d}f/\mathrm{d}t = \dot{f}$ and
$\mathrm{d}f/\mathrm{d}\tau = \check{f}$.

Unless otherwise specified, we will assume that the frame vector
associated with coordinate $x^0$ is unitary and normal to all the
others, that is $g_0 = \sigma_0$ and $g_{0 i} = 0$. Recalling from
Eq.\ (\ref{eq:4dometric}), these conditions allow the definition of
4DO space with metric tensor $g_{ij}$. Although we could try a more
general approach, we would loose the possibility of interpreting
time as a line element and this, as we shall see, provides very
interesting and novel interpretations of physics' equations. In many
cases it is also true that $g_4$ is normal to the other frame
vectors and we have seen that in those cases we can make metric
conversions between GTR and 4DO; as we shall see, electromagnetism
requires a non-normal $g_4$ and so we leave this possibility open.

For the moment we will concentrate on isotropic space, characterized
by orthogonal refractive index vectors $g_i$ whose norm can change
with coordinates but is the same for all vectors. Normally we relax
this condition by accepting that the three $g_m$ must have equal
norm but $g_4$ can be different. The reason for this relaxed
isotropy is found in the parallel we make with physics by assigning
dimensions $1$ to $3$ to physical space. Isotropy in a physical
sense need only be concerned with these dimensions and ignores what
happens with dimension 4. We will therefore characterize an
isotropic space by the refractive index frame $g_0 = \sigma_0$, $g_m
= n_r \sigma_m$, $g_4 = n_4 \sigma_4$. Indeed we could also accept a
non-orthogonal $g_4$ within the relaxed isotropy concept but we will
not do so for the moment.

Equation (\ref{eq:4dometric}) can now be written in terms of the
isotropic refractive indices as
\begin{equation}
    \mathrm{d}t^2 = (n_r)^2 \sum_m (\mathrm{d}x^m)^2 + (n_4 \mathrm{d}\tau)^2.
\end{equation}
Spherically symmetric static metrics play a special role; this means
that the refractive index can be expressed as functions of $r$ if we
adopt spherical coordinates. The previous equation then becomes
\begin{equation}
\label{eq:spher4do}
    \mathrm{d}t^2 = (n_r)^2 \left[\mathrm{d}r^2 + r^2 (\mathrm{d}\theta^2
    + \sin^2 \theta \mathrm{d}\varphi^2)\right]+ (n_4 \mathrm{d}\tau)^2.
\end{equation}
Since we have $g_4$ normal to the other vectors we can apply metric
conversion and write the equivalent quadratic form for GTR
\begin{equation}
\label{eq:sphergr}
    \mathrm{d}\tau^2 = \left(\frac{\mathrm{d}t}{n_4}\right)^2
    -  \left(\frac{n_r}{n_4}\right)^2  \left[\mathrm{d}r^2 + r^2 (\mathrm{d}\theta^2
    + \sin^2 \theta \mathrm{d}\varphi^2)\right].
\end{equation}

In the case of a central mass, we can examine how the Schwarzschild
metric in GTR can be transposed to 4DO. The usual form of the metric
is
\begin{eqnarray}
    \mathrm{d}\tau^2 &=& \left(1-\frac{2M}{\chi} \right)\mathrm{d}t^2
    -\left(1-\frac{2M}{\chi} \right)^{-1}\mathrm{d}\chi^2 - \nonumber \\
    && - \chi^2
    \left(\mathrm{d}\theta^2 + \sin^2 \theta \mathrm{d}\varphi^2 \right);
    \label{eq:schwarz}
\end{eqnarray}
where $M$ is the spherical mass and $\chi$ is the radial coordinate,
not the distance to the centre of the mass. This form is
non-isotropic but a change of coordinates can be made that returns
the expression to isotropic form (see \citet[section
14.7]{Inverno96}):
\begin{equation}
    r=\left(\chi-M+\sqrt{\chi^2-2M \chi}\right)/2;
\end{equation}
and the new form of the metric is
\begin{equation}
    \label{eq:isotropic}
    \mathrm{d}\tau^2 =
    \left(\frac{\displaystyle 1-\frac{M}{2r}}{\displaystyle 1+\frac{M}{2r}}\right)^2
    \mathrm{d}t^2 -
    \left(1+ \frac{M}{2r}\right)^4  \left[ \mathrm{d}r^2 - r^2 \left(\mathrm{d}\theta^2
    + \sin^2 \theta \mathrm{d}\varphi^2 \right) \right].
\end{equation}
From this equation we immediately define two coefficients, which are
called refractive index coefficients,
\begin{equation}
    \label{eq:nrn4}
    n_4 = \frac{\displaystyle 1+ \frac{M}{2r}}{\displaystyle 1-\frac{M}{2r}},
    ~~~~ n_r = \frac{\left(\displaystyle 1 + \frac{M}{2r}\right)^3}
    {\displaystyle 1-\frac{M}{2r}}.
\end{equation}
These refractive indices provide a 4DO Euclidean space equivalent to
Schwarzschild metric, allowing 4DO to be used as an alternative to
GTR. Recalling that we derived trajectories from solutions
(\ref{eq:psidef}) of a 4-dimensional wave equation
(\ref{eq:curvwave2}), it becomes clear that orbits can also be seen
as 4-dimensional guided waves by what could be described as a
4-dimensional optical fibre. Modes are to be expected in these
waveguides and we shall say something about them later on.

\section{Fermat's principle in 4 dimensions}

Fermat's principle applies to optics and states that the path
followed by a light ray is the one that makes the travel time an
extremum; usually it is the path that minimizes the time but in some
cases a ray can follow a path of maximum or stationary time. These
solutions are usually unstable, so one takes the view that light
must follow the quickest path. In Eq.\ (\ref{eq:4dometric}) we have
defined a time interval associated with a 4-dimensional elementary
displacement, which allows us to determine, by integration, a travel
time associated with displacements of any size along a given
4-dimensional path. We can then extend Fermat's principle to 4D and
impose an extremum requirement in order to select a privileged path
between any two 4D points. Taking the square root to Eq.\
(\ref{eq:4dometric})
\begin{equation}
    \mathrm{d}t = \sqrt{-\frac{g_{ij}}{g_{00}}\, \mathrm{d}x^i
    \mathrm{d}x^j}.
\end{equation}
Integrating between two points $P_1$ and $P_2$
\begin{equation}
    t = \int_{P_1}^{P_2} \sqrt{-\frac{g_{ij}}{g_{00}}\, \mathrm{d}x^i
    \mathrm{d}x^j} = \int_{P_1}^{P_2} \sqrt{-\frac{g_{ij}}{g_{00}}\, \dot{x}^i
    \dot{x}^j}\, \mathrm{d}t.
\end{equation}
In order to evaluate the previous integral one must know the
particular path linking the points by defining functions $x^i(t)$,
allowing the replacement $\mathrm{d}x^i = \dot{x}^i \mathrm{d}t$. At
this stage it is useful to define a Lagrangian
\begin{equation}
\label{eq:lagrangian}
    L =  -\frac{g_{ij}}{2 g_{00}}\, \dot{x}^i
    \dot{x}^j.
\end{equation}
The time integral can then be written
\begin{equation}
\label{eq:time}
    t = \int_{P_1}^{P_2} \sqrt{2 L}\, \mathrm{d}t.
\end{equation}

Time has to remain stationary against any small change of path;
therefore we envisage a slightly distorted path defined by functions
$x^i(t) + \varepsilon \chi^i(t)$, where $\varepsilon$ is arbitrarily
small and $ \chi^i(t)$ are functions that specify distortion. Since
the distortion must not affect the end points, the distortion
functions must vanish at those points. The time integral will now be
a function of $\varepsilon$ and we require that
\begin{equation}
    \left.
    \frac{\mathrm{d}t(\varepsilon)}{\mathrm{d}\varepsilon}\,\right|_{\varepsilon
    =0} = 0.
\end{equation}
Now, the Lagrangian (\ref{eq:lagrangian}) is a function of $x^i$,
through $g_{\alpha \beta}$ and also an explicit function of
$\dot{x}^i$. Allowing for a path change, through $\varepsilon$ makes
$t$ in Eq.\ (\ref{eq:time}) a function of $\varepsilon$
\begin{equation}
    t(\varepsilon) = \int_{P_1}^{P_2} \sqrt{2 L(x^i + \varepsilon
    \chi^i + \dot{x}^i + \varepsilon \dot{\chi}^i)}\, \mathrm{d}t.
\end{equation}
This can now be derived with respect to $\varepsilon$
\begin{equation}
\label{eq:vareq}
    \left.
    \frac{\mathrm{d}t(\varepsilon)}{\mathrm{d}\varepsilon}\,\right|_{\varepsilon
    =0} = \left[
    \int_{P_1}^{P_2} \frac{1}{\sqrt{2L}}\right.  \left. \left(\frac{\partial
    L}{\partial \dot{x}^i}\dot{\chi}^i + \frac{\partial L}{\partial
    x^i}\chi^i \right)\mathrm{d}t \right]_{\varepsilon
    =0}.
\end{equation}
Note that the first term on the rhs can be written
\begin{equation}
    \int_{P_1}^{P_2} \frac{1}{\sqrt{2L}}\frac{\partial
    L}{\partial \dot{x}^i}\dot{\chi}^i \mathrm{d}t =
    \int_{P_1}^{P_2} \frac{\partial (\sqrt{2L})}{\partial \dot{x^i}}
    \dot{\chi}^i \mathrm{d}t.
\end{equation}
This can be integrated by parts
\begin{equation}
    \int_{P_1}^{P_2} \frac{\partial (\sqrt{2L})}{\partial \dot{x^i}}
    \dot{\chi}^i \mathrm{d}t = \left[ \frac{\partial (\sqrt{2L})}
    {\partial \dot{x^i}} \chi^i\right]_{P_1}^{P_2} -  \int_{P_1}^{P_2}
    \frac{\mathrm{d}}{ \mathrm{d} t} \left(\frac{\partial (\sqrt{2L})}
    {\partial \dot{x^i}}\right)\chi^i\mathrm{d}t.
\end{equation}
The first term on the second member is zero because $\chi^i$
vanishes for the end points; replacing in Eq.\ (\ref{eq:vareq})
\begin{equation}
        \left.
    \frac{\mathrm{d}t(\varepsilon)}{\mathrm{d}\varepsilon}\,\right|_{\varepsilon
    =0} = \frac{1}{\sqrt{2}} \int_{P_1}^{P_2}
    \left[\frac{ \mathrm{d}}{ \mathrm{d} t} \left(- \frac{1}{\sqrt{L}}
    \frac{\partial L}{\partial \dot{x}^i}\right) \right.  + \left.
    \frac{1}{\sqrt{L}} \frac{\partial L}{\partial x^i} \right]
    \chi^i\mathrm{d}t.
\end{equation}
The rhs must be zero for arbitrary distortion functions $\chi^i$, so
we conclude that the following set of four simultaneous equations
must be verified
\begin{equation}
\label{eq:euler}
    \frac{ \mathrm{d}}{ \mathrm{d} t} \left( \frac{1}{\sqrt{L}}
    \frac{\partial L}{\partial \dot{x}^i}\right)
    =  \frac{1}{\sqrt{L}} \frac{\partial L}{\partial x^i};
\end{equation}
these are called the Euler-Lagrange equations.

Consideration of Eqs.\ (\ref{eq:velocity}) and (\ref{eq:grmetric})
allows us to conclude that the Lagrangian defined by
(\ref{eq:lagrangian}) can also be written as $L = v^2/2$ and must
always equal $1/2$. From the Lagrangian one defines immediately the
conjugate momenta
\begin{equation}
    v_i = \frac{\partial L}{\partial \dot{x}^i} = \frac{-g_{i j}}{g_{00}} \dot{x}^j.
\end{equation}
Notice the use of the lower index ($v_i$) to represent momenta while
velocity components have an upper index ($v^i$). The conjugate
momenta are the components of the conjugate momentum vector
\begin{equation}
    v = \frac{g^i v_i}{\sqrt{-g_{00}}}
\end{equation}
and from Eq.\ (\ref{eq:recframe})
\begin{equation}
    \label{eq:momentvel}
    \sqrt{-g_{00}} v = g^i v_i = g^i g_{i j} \dot{x}^j = g_j \dot{x}^j.
\end{equation}
The conjugate momentum and velocity are the same but their
components are referred to the reciprocal and refractive index
frames, respectively.\footnote{In most cases $g_{00} = -1$, the
velocity can be conveniently written $v =g_i \dot{x}^i$ and
conjugate momenta $v_i = g_{ij} \dot{x}^j$.} Notice also that by
virtue of Eq.\ (\ref{eq:pnull}) it is also
\begin{equation}
\label{eq:4dveloc}
  v_i = \frac{p_i}{p_0}\, .
\end{equation}

The Euler-Lagrange equations (\ref{eq:euler}) can now be given a
simpler form
\begin{equation}
\label{eq:eulersimp}
    \dot{v}_i = \partial_i L.
\end{equation}
This set of four equations defines trajectories of minimum time in
4DO space as long as the frame vectors $g_\alpha$ are known
everywhere, independently of the fact that they may or may not be
referred to the orthonormed frame via a refractive index. By
definition these trajectories are the geodesics of 4DO space,
spanned by frame vectors $g_i/\sqrt{-g_{00}}$, with metric tensor
$-g_{ij}/g_{00}$.

Following an exactly similar procedure we can find trajectories
which extremize proper time, defined by taking the positive square
root of Eq.\ (\ref{eq:grmetric}). The Lagrangian is now defined by
\begin{equation}
    \mathcal{L} = -\frac{ 1}{2} \frac{g_{\mu\nu}}{g_{44}}\check{x}^\mu
    \check{x}^\nu.
\end{equation}
Consequently the conjugate momenta are
\begin{equation}
    \nu_\mu = \frac{\partial \mathcal{L}}{\partial \check{x}^\mu}
    = \frac{-g_{\mu\nu}}{g_{44}} \check{x}^\nu.
\end{equation}
From Eq.\ (\ref{eq:pnull}) we have $\nu_\mu = p_\mu/p_4$; the
associated Euler-Lagrange equations are
\begin{equation}
\label{eq:eulergr}
    \check{\nu}_\mu = \partial_\mu \mathcal{L}.
\end{equation}
"These are, by definition, spacelike geodesics of GTR with metric
tensor $-g_{\mu\nu}/g_{44}$ and we have thus defined a method for
one-to-one geodesic mapping between 4DO and spacelike GTR. Recalling
the conditions for this mapping to be valid, all the frame vectors
must be independent of both $t$ and $\tau$ and $g_0$ and $g_4$ must
be normal to the other 3 frame vectors. In tensor terms, all the
$g_{\alpha \beta}$ must be independent from $t$ and $\tau$ and
$g_{0i} = g_{\mu 4} =0$."

\section{The sources of refractive index}

The set of 4 equations (\ref{eq:eulersimp}) defines the geodesics of
4DO space; particularly in cases where there is a refractive index,
it defines trajectories of minimum time but does not tell us
anything about what produces the refractive index in the first
place. Similarly the set of equations (\ref{eq:eulergr}) defines the
geodesics of GTR space without telling us what shapes space. In
order to analyse this question we must return to the general case of
a refractive frame $g_\alpha$ without other impositions besides the
existence of a refractive index.

Considering the momentum vector
\begin{equation}
    p = p_\alpha g^\alpha = p_\alpha {n_\beta}^\alpha \sigma^\beta,
\end{equation}
with ${n_\alpha}^\gamma  {n^\beta}_\gamma = \delta_\alpha^\beta$, we
will now take its time derivative. Using Eq.\ (\ref{eq:timeder})
\begin{equation}
\label{eq:dotp}
    \dot{p} = \dot{x} \dprod (\mathrm{D}p) = \dot{x} \dprod G.
\end{equation}
By a suitable choice of coordinates we can always have $g^0 =
\sigma^0$. We can then invoke the fact that for an elementary
particle in flat space the momentum vector components can be
associated with the concepts of energy, 3D momentum and rest mass as
$p = E \sigma^0 + \mathbf{p} + m \sigma^4$ (see Sec.\
\ref{dynamics}.) If this consequence is extended to curved space and
to mass distributions, we write $p = E \sigma^0 + \mathbf{p} + m
g^4$, where now $E$ is energy density, $\mathbf{p} = p_m g^m$ is 3D
momentum density and $m$ is mass density. The previous equation then
becomes
\begin{equation}
\label{eq:dynamicst}
    \dot{E} \sigma^0 +\dot{\mathbf{p}} + m \dot{g}^4 = \dot{x} \dprod G.
\end{equation}

When the Laplacian is applied to the momentum vector the result is
still necessarily a vector
\begin{equation}
    \label{eq:current}
    \mathrm{D}^2  p = S.
\end{equation}
Vector $S$ is called the \emph{sources vector} and can be expanded
into 25 terms as
\begin{equation}
\label{eq:sources}
    S = (\mathrm{D}^2 {n^\beta}_\alpha)
    \sigma_\beta p^\alpha =
    {S^\beta}_\alpha \sigma_\beta p^\alpha;
\end{equation}
where $p^\alpha = g^{\alpha \beta} p_\beta$. Tensor
${S^\alpha}_\beta$ contains the coefficients of the sources vector
and we call it the \emph{sources tensor}. The sources tensor
influences the shape of geodesics as we shall see in one
particularly important situation. One important consequence that we
don't pursue here is that by zeroing the sources vector one obtains
the wave equation $\mathrm{D}^2 p =0$, which accepts gravitational
wave solutions.

If $\sigma^0$ is normal to the other frame vectors we can write $p =
E(\sigma^0 + v)$ in the reciprocal frame, with $v$ a unit vector or
$p = E(-\sigma_0 + v)$ in the direct frame. Equation (\ref{eq:dotp})
can then be given the form
\begin{equation}
    \dot{E} (\sigma^0 + v) + E \dot{v} = {\sigma_0 + v} \dprod G.
\end{equation}
Since $G$ can have scalar and bivector components, the scalar part
must be responsible for the energy change, while the bivector part
rotates the velocity $v$. The bivector part of $G$ is generated by
$\mathrm{D} \wprod p$, which allows a simplification of the previous
equation to
\begin{equation}
    \dot{v} = v \dprod (\mathrm{D} \wprod v),
\end{equation}
if the frame vectors are independent of $t$. This equation is
exactly equivalent to the set of Euler-Lagrange equations
(\ref{eq:eulersimp}) but it was derived in a way which tells us when
to expect geodesic movement or free fall.

We will now investigate spherically symmetric solutions in isotropic
conditions defined by Eq.\ (\ref{eq:spher4do}); this means that the
refractive index can be expressed as functions of $r$. The vector
derivative in spherical coordinates is of course
\begin{equation}
    \mathrm{D} = \frac{1}{n_r}\, \left(\sigma_r \partial_r + \frac{1}{r}\,
    \sigma_\theta \partial_\theta + \frac{1}{r \sin \theta}\, \sigma_\varphi
    \partial_\varphi \right) - \sigma_t \partial_t   + \frac{1}{n_4}\, \sigma_\tau \partial_\tau.
\end{equation}
The Laplacian is the inner product of $\mathrm{D}$ with itself but
the frame vectors' derivatives must be considered; all the
derivatives with respect to $t$, $r$ and $\tau$ are zero and the
non-zero ones are
\begin{equation}
    \begin{array}{ll}
      \partial_\theta \sigma_r = \sigma_\theta, &
      \partial_\varphi \sigma_r = \sin \theta \sigma_\varphi, \\
      \partial_\theta \sigma_\theta = -\sigma_r, &
      \partial_\varphi \sigma_\theta = \cos \theta \sigma_\varphi, \\
      \partial_\theta \sigma_\varphi = 0, &
      \partial_\varphi \sigma_\varphi = -\sin \theta\, \sigma_r - \cos \theta\, \sigma_\theta.
    \end{array}
\end{equation}
After evaluation the curved Laplacian becomes
\begin{eqnarray}
    \label{eq:laplacradial}
    \mathrm{D}^2 &=& \frac{1}{(n_r)^2}\, \left(\partial_{rr} + \frac{2}{r}\, \partial_r -
    \frac{n'_r}{n_r}\, \partial_r  + \frac{1}{r^2}\, \partial_{\theta \theta}
     \right . + \nonumber \\
    && \left .
    +\frac{\cot \theta}{r^2}\, \partial_\theta
    + \frac{\csc^2 \theta}{r^2}\, \partial_{\varphi \varphi} \right)
    - \partial_{tt} + \frac{1}{(n_4)^2}\, \partial_{\tau \tau}.
\end{eqnarray}

The search for solutions of Eq.\ (\ref{eq:current}) must necessarily
start with vanishing second member, a zero sources situation, which
one would implicitly assign to vacuum; this is a wrong assumption as
we will show. Zeroing the second member implies that the Laplacian
of both $n_r$ and $n_4$ must be zero; considering that they are
functions of $r$ we get the following equation for $n_r$
\begin{equation}
    n^{''}_r + \frac{2 n'_r}{r} - \frac{(n'_r)^2}{n_r} = 0,
\end{equation}
with general solution $n_r = b \exp(a/r)$. It is legitimate to make
$b =1$ because the refractive index must be unity at infinity. Using
this solution in Eq.\ (\ref{eq:laplacradial}) the Laplacian becomes
\begin{eqnarray}
    \mathrm{D}^2 &=& \mathrm{e}^{-a/r}\left(\partial_{rr} + \frac{2}{r}\,
    \partial_r
     + \frac{a }{r^2}\, \partial_r + \frac{1}{r^2}\,\, \partial_{\theta \theta}
     \right . + \nonumber \\
    && \left .
    +\frac{\cot \theta}{r^2} \, \partial_\theta
    + \frac{\csc^2 \theta}{r^2}\, \partial_{\varphi \varphi}\right)
     - \partial_{tt} + \frac{1}{(n_4)^2}\, \partial_{\tau \tau};
\end{eqnarray}
which produces the solution $n_4 = n_r $. So space must be truly
isotropic and not relaxed isotropic as we had allowed. The solution
we have found for the refractive index components in isotropic space
can correctly model Newton dynamics, which led the author to adhere
to it for some time \cite{Almeida01:4}. However if inserted into
Eq.\ (\ref{eq:grmetric}) this solution produces a GTR metric which
is verifiably in disagreement with observations; consequently it has
purely geometric significance.

The inadequacy of the isotropic solution found above for
relativistic predictions deserves some thought, so that we can
search for solutions guided by the results that are expected to have
physical significance. In the physical world we are never in a
situation of zero sources because the shape of space or the
existence of a refractive index must always be tested with a test
particle. A test particle is an abstraction corresponding to a point
mass considered so small as to have no influence on the shape of
space; in reality a point particle is a black hole in GTR, although
this fact is always overlooked; one wonders how a black hole is
postulated not to influence space geometry. A test particle must be
seen as source of refractive index itself and its influence on the
shape of space should not be neglected in any circumstances. If this
is the case the solutions for vanishing sources vector may have only
geometric meaning, with no connection to physical reality.

The question is then what should go into the second member of Eq.\
(\ref{eq:current}) in order to find physically meaningful solutions.
If we are testing gravity we must assume some mass density to suffer
gravitational influence; this is what is usually designated as
non-interacting dust, meaning that some continuous distribution of
non-interacting particles follows the geodesics of space. Mass
density is expected to be associated with ${S^4}_4$; on the other
hand we are assuming that this mass density is very small and so we
use flat space Laplacian to evaluate it. We consequently make an
\emph{ad hoc} proposal for the sources vector in the second member
of Eq.\ (\ref{eq:current})
\begin{equation}
    \label{eq:statpart}
    S = -\nabla^2 n_4 \sigma_4.
\end{equation}
Equation (\ref{eq:current}) becomes
\begin{equation}
    \label{eq:gravitation}
    \mathrm{D}^2 \dot{x} = -\nabla^2 n_4 \sigma_4;
\end{equation}
as a result the equation for $n_r$ remains unchanged but the
equation for $n_4$ becomes
\begin{equation}
    n^{''}_4 + \frac{2 n'_4}{r} - \frac{n'_r n'_4}{n_r}
    = - n^{''}_4 + \frac{2 n'_4}{r}.
\end{equation}

When $n_r$ is given the exponential form found above, the solution
is $n_4 = \sqrt{n_r}$. This can now be entered into Eq.\
(\ref{eq:grmetric}) and the coefficients can be expanded in series
and compared to Schwarzschild's for the determination of parameter
$a$. The final solution, for a stationary mass $M$ is
\begin{equation}
    \label{eq:refind}
    n_r = \mathrm{e}^{2M/r},~~~~n_4 = \mathrm{e}^{M/r}.
\end{equation}
The equivalent GTR space is characterized by the quadratic form
\begin{equation}
    \mathrm{d}\tau^2 = \mathrm{e}^{-2M/r}\mathrm{d}t^2 - \mathrm{e}^{2M/r}\sum_m
    (\mathrm{d}x^m)^2.
\end{equation}
Expanding in series of $M/r$ the coefficients of this metric one
would find that the lower order terms are exactly the same as for
Schwarzschild's and so the predictions of the metrics are
indistinguishable for small values of the expansion variable.
\citet{Montanus01} arrives at the same solutions with a different
reasoning; Yilmaz was probably the first author to propose this
metric \cite{Yilmaz58,Yilmaz71,Ibison05}.

Equation (\ref{eq:gravitation}) can be interpreted in physical terms
as containing the essence of gravitation. When solved for
spherically symmetric solutions, as we have done, the first member
provides the definition of a stationary gravitational mass as the
factor $M$ appearing in the exponent and the second member defines
inertial mass as $\nabla^2 n_4$. Gravitational mass is defined with
recourse to some particle which undergoes gravitational influence
and is animated with velocity $v$ and inertial mass cannot be
defined without some field $n_4$ acting upon it. Complete
investigation of the sources tensor elements and their relation to
physical quantities is not yet done; it is believed that  16 terms
of this tensor have strong links with homologous elements of stress
tensor in GTR, while the others are related to electromagnetic
field.

\section{Electromagnetism in 5D spacetime \label{EM}}
Maxwell's equations can easily be written in the form of Eq.\
(\ref{eq:current}) if we don't impose the condition that $g_4$
should remain normal the other frame vectors; as we have seen in
section \ref{dynamics} this has the consequence that there will be
no GTR equivalent to the equations formulated in 4DO.

We will consider the non-orthonormed reciprocal frame defined by
\begin{equation}
    g^\mu = \sigma^\mu,~~~~ g^4 = \frac{q}{m}\, A_\mu \sigma^\mu +
    \sigma^4;
\end{equation}
where $q$ and $m$ are charge and mass densities, respectively, and
$A = A_\mu \sigma^\mu$ is the electromagnetic vector potential,
assumed to be a function of coordinates $t$ and $x^m$ but
independent of $\tau$. The associated direct frame has vectors
\begin{equation}
    g_\mu = \sigma_\mu - \frac{q}{m}\, A_\mu \sigma_4,~~~~g_4 =
    \sigma_4;
\end{equation}
and one can easily verify that Eq.\ (\ref{eq:recframe}) is obeyed.
The momentum vector in the reciprocal frame is $p = E \sigma^0 + p_m
\sigma^m + q A_\mu \sigma^\mu + m \sigma^4$ and $G$ in the second
member of Eq.\ (\ref{eq:dotp}) is $G = q \mathrm{D} A$. We will
assume $\mathrm{D} \dprod A$ to be zero, as one usually does in
electromagnetism; also $\mathrm{D}$ can be replaced by $\pre{\mu}
\nabla$ because the vector potential does not depend on $\tau$. It
is convenient to define the Faraday bivector $F =\, \pre{\mu} \nabla
A$, similarly to what is done in Ref.\ \cite{Doran03}; the dynamics
equation then becomes
\begin{equation}
    \dot{\mathbf{p}} +  q \dot{A} = q \dot{x} \dprod F;
\end{equation}
and rearranging
\begin{equation}
    \dot{\mathbf{p}} = q \dot{x} \dprod F - q \dot{A}.
\end{equation}
The first term in the second member is the Lorentz force and the
second term is due to the radiation of an accelerated charge.

Recalling the wave displacement vector Eq.\ (\ref{eq:wavedisp}) we
have now
\begin{equation}
    \mathrm{d}{x} =  \sigma_\alpha
    \mathrm{d}{x}^\alpha - \frac{q}{m}\,  A_\mu \sigma_4 \mathrm{d}{x}^\mu.
\end{equation}
This corresponds to a refractive index tensor whose non-zero terms
are
\begin{equation}
    {n^\alpha}_\alpha = 1,~~~~ {n^4}_\mu = -\frac{q}{m}\, A_\mu.
\end{equation}

According to Eq.\ (\ref{eq:sources}) the sources tensor has all
terms null except for the following
\begin{equation}
\label{eq:emsources}
    {S^4}_\mu = -\frac{q}{m}\, \mathrm{D}^2 A_\mu;
\end{equation}
where $\mathrm{D}$ is the covariant derivative given by
\begin{equation}
\label{eq:EMD}
    \mathrm{D} = g^\alpha \partial_\alpha = \sigma^\mu \partial_\mu +
    (\sigma^4 + \frac{q}{m}\, A_\mu \sigma^\mu) \partial_4.
\end{equation}
We can then define the current vector $J$ verifying
\begin{equation}
\label{eq:Maxwell}
    \pre{\mu}\nabla^2 A =\, \pre{\mu} \nabla F = J,
\end{equation}
where
\begin{equation}
    J = -\frac{m}{q}\, {S^4}_\mu \sigma^\mu.
\end{equation}
Please refer to \cite[Chap.\ 7]{Doran03} or to \cite[Part
2]{Lasenby99} to see how these equations generate classical
electromagnetism.

In free space we make $J = 0$ and Eq.\ (\ref{eq:Maxwell}) accepts
plane wave solutions for $F$ which are of course electromagnetic
waves. Notice that these solutions propagate in directions normal to
proper time, which is perfectly consistent with the classical
relativistic formulation.

The Dirac equation for a free particle has been derived from the
5-dimensional monogenic condition in Sec.\ \ref{dynamics} but we are
now in position to include the effects of an EM field. Because we
are working in geometric algebra, our quantum mechanics equations
will inherit that character but the isomorphism between the
geometric algebra of 5D spacetime, $G_{4,1}$, and complex algebra of
$4 \ast 4$ matrices, $M(4,C)$, ensures that they can be translated
into the more usual Dirac matrix formalism. Electrodynamics can now
be implemented in the the same way used in Sec.\ \ref{EM} to
implement classical electromagnetism. The monogenic condition must
now be established with the covariant derivative given by Eq.\
(\ref{eq:EMD})
\begin{equation}
    \sigma^\mu \partial_\mu \psi + \left(\sigma^4 + \frac{q}{m}\, A_\mu \sigma^\mu \right)
    \partial_4 \psi = 0.
\end{equation}
Multiplying on the left by $\sigma^4$ and taking $\partial_4 \psi =
\mathrm{i}m \psi$
\begin{equation}
    \left[\gamma^\mu (\partial_\mu + \mathrm{i} q A_\mu) + \mathrm{i}m \right]
    \psi = 0.
\end{equation}
This equation can be compared to what is found in any quantum
mechanics textbook..

It is now adequate to say a few words about quantization, which is
inherent to 5D monogenic functions. We have already seen that these
functions are 4-dimensional waves, that is, they have 3-dimensional
wavefronts normal to the direction of propagation. Whenever the
refractive index distribution traps one of these waves a
4-dimensional waveguide is produced, which has its own allowed
propagating modes. In the particular case of a central potential, be
it an atom's or a galaxy's nucleus, we expect spherical harmonic
modes, which produce the well known electron orbitals in the atom
and have unknown manifestations in a galaxy.

\section{Hyperspherical coordinates \label{hyperspherical}}
Deriving physical equations and predictions from purely geometrical
equations is an exercise whose success depends on the correct
assignment of coordinates to physical entities; the same space will
produce different predictions if different options are taken for
coordinate assignment. In the previous sections we assumed that
empty space could be modelled by an assignment of time, three
spatial directions and proper time to five orthogonal directions in
5D spacetime. We are now going to experiment with a different
assignment of flat space coordinates, which will explore the
possibility that physics and the Universe have an inbuilt
hyperspherical symmetry. The exercise consists on assigning
coordinate $x^4 = \tau$ to the radius of an hypersphere and the
three $x^m$ coordinates to distances measured on the hypersphere
surface; time, $x^0$, will still be measured along a direction
normal to all others. If the hypersphere radius is very large we
will not be able to notice the curvature on everyday phenomena, in
the same way as everyday displacements on Earth don't seem curved to
us. The Universe as a whole will manifest the consequences of its
hyperspherical symmetry; using the Earth as a 3-dimensional analogue
of an hyperspherical Universe, although our everyday life is greatly
unaffected by Earth's curvature the atmosphere senses this curvature
and shows manifestations of it in winds and climate. What we propose
here is an exercise consisting of an arbitrary assignment between
coordinates and physical entities; the validity of such exercise can
only be judged by the predictions it allows and how well they
conform with observations.

Hyperspherical coordinates are characterized by one distance
coordinate, $\tau$ and three angles $\rho, \theta, \varphi$;
following the usual procedure we will associate with these
coordinates the frame vectors $\{\sigma_\tau, \sigma_\rho,
\sigma_\theta, \sigma_\varphi\}$. The position vector for one point
in 5D space is quite simply
\begin{equation}
    x = t \sigma_t + \tau \sigma_\tau.
\end{equation}
In order to write an elementary displacement $\mathrm{d}x$ we must
consider the rotation of frame vectors, but we don't need to think
hard about it because we can extend what is known from ordinary
spherical coordinates.
\begin{equation}
    \label{eq:hyperdisplacement}
    \mathrm{d}x = \sigma_0 \mathrm{d}t + \sigma_4 \mathrm{d}\tau + \tau \sigma_\rho
    \mathrm{d}\rho + \tau \sin \rho \sigma_\theta \mathrm{d}\theta +
    \tau \sin \rho \sin \theta \sigma_\varphi \mathrm{d}\varphi.
\end{equation}
Just as before, we consider only null displacements to obtain time
intervals;
\begin{equation}
    \mathrm{d}t^2 = \mathrm{d}\tau^2 + \tau^2 \left[\mathrm{d}\rho^2 +
    \sin^2 \rho \left(\mathrm{d}\theta^2 +
    \sin^2 \theta \mathrm{d}\varphi^2 \right)\right].
\end{equation}
The velocity vector, $v = \dot{x} - \sigma_0$, can be immediately
obtained from the displacement vector dividing by $\mathrm{d}t$
\begin{equation}
    \label{eq:velhspher}
    v = \sigma_0 \dot{\tau} + \tau \sigma_\rho
    \dot{\rho} + \tau \sin \rho \sigma_\theta \dot{\theta} +
    \tau \sin \rho \sin \theta \sigma_\varphi \dot{\varphi}.
\end{equation}

Geodesics of flat space are naturally straight lines, no matter
which coordinate system we use, however it is useful to derive
geodesic equations from a Lagrangian of the form
\eqref{eq:lagrangian}; in hyperspherical coordinates the Lagrangian
becomes
\begin{equation}
   2 L = v^2 = \dot{\tau}^2 + \tau^2 \left[\dot{\rho}^2 + \sin^2
   \rho \left(\dot{\theta}^2 + \sin^2 \theta \dot{\varphi}^2 \right)
   \right].
\end{equation}
Because de Lagrangian is independent of $\varphi$ we can establish a
conserved quantity
\begin{equation}
    J_\varphi = \tau^2 \sin^2 \rho \sin^2 \theta \dot{\varphi}.
\end{equation}

It may seem strange that any physically meaningful relation can be
derived from the simple coordinate assignment that we have made,
that is, proper time is associated with hypersphere radius and the
three usual space coordinates are assigned to distances on the
hypersphere radius. This unexpected fact results from the
possibility offered by hyperspherical coordinates to explore a
symmetry in the Universe that becomes hidden when we use Cartesian
coordinates. In the real world we measure distances between objects,
namely cosmological objects, rather than angles; we have therefore
to define a distance coordinate, which is obviously $r = \tau \rho$.
It does not matter where in the Universe we place the origin for $r$
and we find it convenient to place ourselves on the origin.

Radial velocities $\dot{r}$ measure movement in a radial direction
from our observation point; we are particularly interested in this
type of movement in order to find a link to the Hubble relation.
Applying the chain rule and then replacing $\rho$
\begin{equation}
    \dot{r} = \rho \dot{\tau} + \dot{\rho} \tau =
    \frac{\dot{\tau}}{\tau}\, r + \dot{\rho} \tau.
\end{equation}
We expect objects that have not suffered any interaction to move
along $\sigma_\tau$; from \eqref{eq:velhspher} we see that this
implies $\dot{\rho} = \dot{\theta} = \dot{\varphi}=0$ and then
$\dot{\tau}$ becomes unity. Replacing in the equation above and
rearranging
\begin{equation}
    \label{eq:firsthubble}
    \frac{\dot{r}}{r} = \frac{1}{\tau}.
\end{equation}
What this equation tells us is exactly what is expressed by the
Hubble relation. The value of $\tau$ can be taken as constant for
any given observation because the distance information is carried by
photons and these preserve proper time, as we have seen in our
discussion about electromagnetic waves.\footnote{In order to
preserve proper time photons must travel on the hypersphere surface
and thus don't follow geodesics.} The first member of the equation
is the definition of the Hubble parameter and we can then write $H =
1/\tau$. In this way we find the physical meaning of coordinate
$\tau$ as being the Universe's age.

Underlying the present discussion there is an assumption a preferred
frame where stillness means moving along $\sigma_\tau$; there is no
question of equivalent inertial frames here. This preferred frame is
obviously attached to the observable still objects in the Universe
which are galaxy clusters, as much as we can tell. This is far from
the orthodox point of view, because galaxy clusters are seen as
moving relative to each other and so cannot possible define a fixed
frame. But in our formulation still objects move in straight lines
along the proper time direction and keep their angular separations
constant; this is naturally perceived as increasing mutual
distances. If there is any relation between our formulation and an
ether it must be found in the fact that movement has an absolute
meaning, so it is defined relative to something that is fixed; we
call the fixed reference a preferred frame while other authors call
it ether.

How does the use of hyperspherical coordinates affect dynamics in
our laboratory experiments? We would like to know if these
coordinates need only be considered in problems of cosmological
scale or, on the contrary, there are implications for everyday
experiments. The answer implies rewriting
\eqref{eq:hyperdisplacement} with distance rather than angle
coordinates; replacing $\rho$,
\begin{equation}
    \label{eq:dxhyper}
    \mathrm{d}x = \sigma_0 \mathrm{d}t + \left(\sigma_4 - \frac{r}{\tau}\,
    \sigma_\rho
    \right) \mathrm{d}\tau + \sigma_\rho \mathrm{d}r + r (\sigma_\theta
    \mathrm{d}\theta + \sin \theta \sigma_\varphi
    \mathrm{d}\varphi ).
\end{equation}
Evaluating time intervals from the null displacement condition, as
before
\begin{equation}
    \mathrm{d}t^2 = \left[1+ \left(\frac{r}{\tau}\right)^2 \right]\mathrm{d}\tau^2
    - 2 \frac{r}{\tau}\, \mathrm{d}\tau
    \mathrm{d}r + \mathrm{d}r^2 + r^2 (\mathrm{d}\theta^2 + \sin^2
    \theta \mathrm{d}\varphi^2).
\end{equation}
This would be a version of \eqref{eq:twospaces} in spherical
coordinates, were it not for the extra terms with powers of $r/\tau$
in the second member. The coefficient $r/\tau$ implies a comparison
between the distance from the object to the observer and the size of
the Universe; remember that $\tau$ is both time and distance in
non-dimensional units. We can say that ordinary special relativity
will apply for objects which are near us, but distant objects will
show in their movement an effect of the Universe's hyperspherical
nature.

With Eqs.\ (\ref{eq:refind}) we have established the refractive
indices $n_r$  and $n_4$  to account for the dynamics near a massive
sphere using Cartesian coordinates; since this is frequently applied
on a cosmological scale, we must find out how the dynamics is
modified by the use of hyperspherical coordinates. Using the
refractive indices and hyperspherical coordinates, noting that $n_r
= n_4^2$, Eq.\ (\ref{eq:4dometric}) becomes
\begin{equation}
    \mathrm{d}t^2 = n_4^2 \mathrm{d}\tau^2 + n_4^4 \tau^2
    \mathrm{d}\rho^2.
\end{equation}
Dividing both members by $\mathrm{d}t^2$ and reversing the equation
\begin{equation}
    \label{eq:radvel}
    n_4^2 \dot{\tau}^2 + n_4^4 \tau^2 \dot{\rho}^2 = 1;
\end{equation}
and replacing $\tau \dot{\rho}$ by $\dot{r}- r \dot{\tau}/\tau$
\begin{equation}
    n_4^2 \dot{\tau}^2 + n_4^4 \left[\dot{r}^2 + \left(
    \frac{\dot{\tau}}{\tau} \right)^2 r^2 - 2 \dot{\tau} \dot{r}
    \frac{r}{\tau} \right] = 1.
\end{equation}
Dividing both members by $n_4^4 r^2$ and rearranging results in the
equation
\begin{equation}
    \label{eq:dotrr}
    \left(\frac{\dot{r}}{r} \right)^2 =  \left(
    \frac{1}{n_4^4} - \frac{\dot{\tau}^2}{n_4^2}
    \right)\frac{1}{r^2} -\left(\frac{\dot{\tau}}{\tau} \right)^2
     + 2 \frac{\dot{\tau} \dot{r}}{\tau r}.
\end{equation}
As a further step we take the refractive index coefficients from
Schwarzschild's metric (\ref{eq:nrn4}) or those of from the
exponential metric (\ref{eq:refind}) and expand the second member in
series of $M/r$ taking only the two first terms.
\begin{equation}
    \label{eq:newfriedman}
    \left(\frac{\dot{r}}{r} \right)^2 \approx \frac{1 - \dot{\tau}^2}{r^2}
    + \frac{(2 \dot{\tau}^2
    -4) M}{r^3} -\left(\frac{\dot{\tau}}{\tau}
    \right)^2 + 2 \frac{\dot{\tau} \dot{r}}{\tau r}.
\end{equation}
The previous equation applies to bodies moving radially under the
influence of mass $M$ located at the origin which is, remember, the
observer's position. For comparison we derive the corresponding
equation in Cartesian coordinates; starting with \eqref{eq:radvel}
it is now
\begin{equation}
    n_4^2 \dot{\tau}^2 + n_4^4 \dot{r}^2 =1;
\end{equation}
dividing by $n_4^4 r^2$ and rearranging
\begin{equation}
    \left(\frac{\dot{r}}{r} \right)^2 = \left(
    \frac{1}{n_4^4} - \frac{\dot{\tau}^2}{n_4^2}
    \right)\frac{1}{r^2} \approx \frac{1 - \dot{\tau}^2}{r^2} + \frac{(2 \dot{\tau}^2
    -4) M}{r^3}.
\end{equation}

If we want to apply these equations to cosmology it is easiest to
follow the approach of Newtonian cosmology, which produces basically
the same results as the relativistic approach but presumes that the
observer is at the centre of the Universe
\cite{Inverno96,Narlikar02}. In order to adopt a relativistic
approach we need equations that replace Einstein's in 4DO. A set of
such was proposed above Eq.\ (\ref{eq:current}) but their
application in cosmology has not yet been tested, so we will have to
defer this more correct approach to future work. The strategy we
will follow here is to consider a general object at distance $r$
from the observer, moving away from the latter under the
gravitational influence of the mass included in a sphere of radius
$r$. If we designate by $\mu$ the average mass density in the
Universe, then mass $M$ in \eqref{eq:newfriedman} is $4 \pi \mu
r^3/3$; this will have to be considered further down.

Friedman equation governs standard cosmology and can be derived both
from Newtonian and relativistic dynamics, with different
consequences in terms of the overall size of the Universe and the
observer's privileged position. From the cited references we write
Friedman equation as
\begin{equation}
    \left(\frac{\dot{r}}{r} \right)^2 = \frac{8 \pi}{3}\, \mu +
    \frac{\Lambda}{3 } - \frac{k }{r^2};
\end{equation}
with $\Lambda$ a cosmological constant and $k$ the curvature
constant; the gravitational constant was not included because it is
unity in non-dimensional units and the equation is written in real,
not comoving, coordinates. In order to compare
\eqref{eq:newfriedman} with Friedman equation there is a problem
with the last term because the Hubble parameter $\dot{r}/r$ does not
appear isolated in the first member; we will find a way to
circumvent the problem later on but first let us look at what
\eqref{eq:newfriedman} tells us when the mass density is zeroed. In
this case $n_4 = 1$ and we find from \eqref{eq:radvel} that
$\dot{\tau}$ is unity, unless $\dot{\rho}$ is non-zero, for which we
can find no reasonable explanation. Replacing $n_4$ and $\dot{\tau}$
with unity in \eqref{eq:newfriedman} we find that $\dot{r}/r =
1/\tau$, confirming what had already been found in
\eqref{eq:firsthubble}. Comparing with Friedman equation, this
corresponds to a flat Universe with a critical mass density $\mu =
\mu_c$; it is immediately obvious that $\mu_c = 3 /(8 \pi \tau^2)$.
Let us not overlook the importance of this conclusion because it
completely removes the need for a critical density if the Universe
is flat; remember this is one of the main reasons to invoke dark
matter in standard cosmology. Notice also that this conclusion does
not depend on a privileged observer, because it is just a
consequence of space symmetry and not of dynamics.

Let us now see what happens when we consider a small mass density;
here we are talking about matter that is observed or measured in
some way but not postulated matter. The matter density that we will
consider is of the order of 1\% of the presently accepted value. It
is therefore just a perturbation of the flat solution that we
described above and the fact that we are presuming a privileged
observer has to be taken just for this perturbation. The first thing
we note when we consider matter density is that $\dot{\tau} <1$,
because there is now a component of the velocity vector along
$\sigma_\rho$. Ideally we should solve the Euler-Lagrange equations
resulting from \eqref{eq:radvel} in order to find $\dot{\tau}$ and
$\dot{\rho}$ but this is a difficult process and we shall carry on
with just a qualitative discussion. Considering that we are
discussing a perturbation it is legitimate to make $\dot{r}/{r}
\approx \dot{\tau}/\tau$ and the two last terms in the second member
of \eqref{eq:newfriedman} can be combined into one single term
$(\dot{\tau}/\tau)^2$, the same as we encountered for the flat
solution, albeit with a numerator slightly smaller than unity. The
first term has now become slightly positive and we can see from
Friedman equation that this corresponds to a negative curvature
constant, $k$, and to an open Universe. Lastly the second term
includes the mass $M$ of a sphere with radius $r$ and can be
simplified to $8 \pi \mu (\dot{\tau}^2 - 2)/3$; this has the effect
of a negative cosmological constant; the combined effect of the two
terms is expected to close the Universe \cite{Martin88,Narlikar02}.
The previous discussion was done in qualitative terms, making use of
several approximations, for which reason we must question some of
the findings and expect that after more detailed examination they
may not be quite as anticipated; in particular there is concern
about the refractive indices used, which were derived in Cartesian
coordinates both by the author and those that preceded him in using
an exponential metric; it may happen that the transposition to
hyperspherical coordinates has not been properly made, with
consequences in the perturbative analysis that was superimposed on
the flat solution. The latter, however, is totally independent of
such concerns and allows us to state that the assumption of
hyperspherical symmetry for the Universe dispenses with dark matter
in accounting for the gross of observed expansion.

Dark matter is also called in cosmology to account for the extremely
high rotation velocities found in spiral galaxies
\cite{Silk97,Rubin78} and we will now take a brief look at how
hyperspherical symmetry can help explain this phenomenon. Galaxy
dynamics is an extremely complex subject, which we do not intend to
explore here due to lack of space but most of all due to lack of
author's competence to approach it with any rigour; we will just
have a very brief outlook at the equation for flat orbits, to notice
that an effect similar to the familiar Coriollis effect on Earth can
arise in an expanding hyperspherical Universe and this could explain
most of the observed velocities on the periphery of galaxies. Let us
recall \eqref{eq:dxhyper}, divide by $\mathrm{d}t$ and invoke null
displacement to obtain the velocity
\begin{equation}
    v = \left(\sigma_4 -
    \frac{r}{\tau}\,
    \sigma_\rho
    \right) \dot{\tau} + \sigma_\rho \dot{r} + r (\sigma_\theta
    \dot{\theta} + \sin \theta \sigma_\varphi
    \dot{\varphi} ).
\end{equation}
If orbits are flat we can make $\theta = \pi/2$ and the equation
simplifies to
\begin{equation}
    \label{eq:rotvel}
    v = \dot{\tau} \sigma_4
    +  \left( \dot{r} -
    \frac{r \dot{\tau}}{\tau} \right) \sigma_\rho + r
    \dot{\varphi} \sigma_\varphi.
\end{equation}
Suppose now that something in the galaxy is pushing outwards
slightly, so that the parenthesis is zero; this happens if
$\dot{r}/r = \dot{\tau}/\tau$ and can be caused by a pressure
gradient, for instance. The result is that \eqref{eq:rotvel} now
accepts solutions with constant $r \dot{\varphi}$, which is exactly
what is observed in many cases; swirls will be maintained by a
radial expansion rate which exactly matches the quotient
$\dot{\tau}/\tau$. In any practical situation $\dot{\tau}$ will be
very near unity and the quotient will be virtually equal to the
Hubble parameter; thus the expansion rate for sustained rotation is
$\dot{r}/r \approx H$. If applied to our neighbour galaxy Andromeda,
with a radial extent of $30~\mathrm{kpc}$, using the Hubble
parameter value of $81~\mathrm{km}\,\mathrm{ s}^{-1}/\mathrm{Mpc}$,
the expansion velocity is about $2.43~\mathrm{km}\,\mathrm{
s}^{-1}$; this is to be compared with the orbital velocity of near
$300~\mathrm{km}\,\mathrm{s}^{-1}$ and probably within the error
margins. An expansion of this sort could be present in many galaxies
and go undetected because it needs only be of the order of 1\% the
orbital velocity.

\section{Symmetries of $G_{4,1}$\index{G(4,1)@$G_{4,1}$} algebra}
In this algebra it is possible to find a maximum of four mutually
annihilating {idempotents}, which generate with $0$ an additive
group of order 16; for a demonstration see \citet[section
17.5]{Lounesto01}. Those {idempotents} can be generated by a choice
of two commuting {basis} elements which square to unity; for the
moment we will use $\sigma_{023}$ and $\sigma_{014}.$ The set of 4
{idempotents} is then given by
\begin{equation}
\begin{split}
    \label{eq:idempotent}
    f_1 = \frac{(1 + \sigma_{023})(1 + \sigma_{014})}{4}, ~~
    f_2 = \frac{(1 + \sigma_{023})(1 - \sigma_{014})}{4}, \\
    f_3 = \frac{(1 - \sigma_{023})(1 - \sigma_{014})}{4}, ~~
    f_4 = \frac{(1 - \sigma_{023})(1 + \sigma_{014})}{4}.
\end{split}
\end{equation}
Using the matrices of Sec.\ \ref{dynamics} to make {matrix}
replacements of $\sigma_{023}$ and $\sigma_{014}$ one can find
{matrix} equivalents to these {idempotents}; those are {matrices}
which have only one non-zero element, located on the diagonal and
with unit value.

$SU(3)$\index{SU(3)@$SU(3)$} {symmetry} can now be demonstrated by
construction of the 8 {generators}
\begin{equation}
\label{eq:su3gen}
\begin{split}
    \lambda_1 &= \sigma_{02} (f_1+f_2)= \frac{\sigma_{3}+\sigma_{02}}{2}, \\
    \lambda_2 &= \sigma_{03} (f_1+f_2)= \frac{-\sigma_{2}+\sigma_{03}}{2}, \\
    \lambda_3 &= f_1 - f_2= \frac{\sigma_{014}-\sigma_{1234}}{2}, \\
    \lambda_4 &= -\sigma_1 (f_2 + f_3)= \frac{-\sigma_{1}-\sigma_{04}}{2}, \\
    \lambda_5 &= -\sigma_4 (f_2 + f_3)= \frac{-\sigma_{4}+\sigma_{01}}{2}, \\
    \lambda_6 &= \sigma_{012} (f_1 + f_3)= \frac{\sigma_{012}+\sigma_{034}}{2}, \\
    \lambda_7 &= -\sigma_{024} (f_1 + f_3)= \frac{\sigma_{013}-\sigma_{024}}{2}, \\
    \lambda_8 &= \frac{f_1 + f_2 - 2 f_3}{\sqrt{3}}= \frac{2 \sigma_{023} + \sigma_{014} +
    \sigma_{1234}}{2 \sqrt{3}}.
\end{split}
\end{equation}
These have the following {matrix} equivalents
\begin{equation}
\label{eq:gelmann}
\begin{split}
    \lambda_1 \equiv  \begin{pmatrix} 0 & 1 & 0 & 0 \\
    1 & 0 & 0 & 0 \\ 0 & 0 & 0 & 0 \\ 0 & 0 & 0 & 0
    \end{pmatrix},~~
    \lambda_2 \equiv & \begin{pmatrix} 0 & -\mathrm{j} & 0 & 0 \\
    \mathrm{j} & 0 & 0 & 0 \\ 0 & 0 & 0 & 0 \\ 0 & 0 & 0 & 0
    \end{pmatrix},~~
    \lambda_3 \equiv  \begin{pmatrix} 1 & 0 & 0 & 0 \\
    0 & -1 & 0 & 0 \\ 0 & 0 & 0 & 0 \\ 0 & 0 & 0 & 0
    \end{pmatrix},\\
    \lambda_4 \equiv  \begin{pmatrix} 0 & 0 & 0 & 0 \\
    0 & 0 & 1 & 0 \\ 0 & 1 & 0 & 0 \\ 0 & 0 & 0 & 0
    \end{pmatrix}  ,  ~~
    \lambda_5 \equiv & \begin{pmatrix} 0 & 0 & 0 & 0 \\
    0 & 0 & -\mathrm{j} & 0 \\ 0 & \mathrm{j} & 0 & 0 \\ 0 & 0 & 0 & 0
    \end{pmatrix}, ~~
    \lambda_6 \equiv  \begin{pmatrix} 0 & 0 & 1 & 0 \\
    0 & 0 & 0 & 0 \\ 1 & 0 & 0 & 0 \\ 0 & 0 & 0 & 0
    \end{pmatrix}  \\
    \lambda_7 \equiv  \begin{pmatrix} 0 & 0 & -\mathrm{j} & 0 \\
    0 & 0 & 0 & 0 \\ 0 & \mathrm{j} & 0 & 0 \\ 0 & 0 & 0 & 0
    \end{pmatrix}, & ~~
    \lambda_8 \equiv \left(1/\sqrt{3}\right) \begin{pmatrix} 1 & 0 & 0 & 0 \\
    0 & 1 & 0 & 0 \\ 0 & 0 & -2 & 0 \\ 0 & 0 & 0 & 0
    \end{pmatrix}  ,
\end{split}
\end{equation}
which reproduce {Gell-Mann} {matrices} in the upper-left $3 \ast 3$
corner \cite{Almeida05:1, Greiner01, Cottingham98}. Since the
algebra is {isomorphic} to complex $4 \ast 4$ {matrix} algebra, one
expects to find higher order symmetries; \citet{Greiner01} show how
one can add 7 additional {generators} to those of
$SU(3)$\index{SU(3)@$SU(3)$} in order to obtain
$SU(4)$\index{SU(4)@$SU(4)$} and the same procedure can be adopted
in {geometric algebra}. We then define the following additional
$SU(4)$\index{SU(4)@$SU(4)$} {generators}
\begin{equation}
\label{eq:su4gen}
\begin{split}
    \lambda_9 &= \sigma_1 (f_1+f_4) = \frac{\sigma_{1}- \sigma_{04}}{2}, \\
    \lambda_{10} &= \sigma_4 (f_1+f_4) = \frac{\sigma_{4}+ \sigma_{01}}{2}, \\
    \lambda_{11} &= -\sigma_{012} (f_2 + f_4) = \frac{-\sigma_{012}- \sigma_{034}}{2}, \\
    \lambda_{12} &= \sigma_{024} (f_2 + f_4) = \frac{\sigma_{013}+ \sigma_{024}}{2}, \\
    \lambda_{13} &= \sigma_{3} (f_3 + f_4) = \frac{\sigma_{3}-\sigma_{02}}{2}, \\
    \lambda_{14} &= \sigma_{2} (f_3 + f_4) = \frac{\sigma_{2}+ \sigma_{03}}{2}, \\
    \lambda_{15} &= \frac{f_1 + f_2 + f_3 - 3 f_4}{\sqrt{6}} =
    \frac{\sigma_{023}
    - \sigma_{014} - \sigma_{1234}}{\sqrt{6}}.
\end{split}
\end{equation}
Once again, making the replacements with Eq.\ (\ref{eq:diracmatrix})
produces the {matrix} equivalent {generators}
\begin{equation}
\label{eq:greiner}
\begin{split}
    \lambda_9 \equiv  \begin{pmatrix} 0 & 0 & 0 & 1 \\
    0 & 0 & 0 & 0 \\ 0 & 0 & 0 & 0 \\ 1 & 0 & 0 & 0
    \end{pmatrix},~~
    \lambda_{10} \equiv & \begin{pmatrix} 0 & 0 & 0 & -\mathrm{j} \\
    0 & 0 & 0 & 0 \\ 0 & 0 & 0 & 0 \\ \mathrm{j} & 0 & 0 & 0
    \end{pmatrix},~~
    \lambda_{11} \equiv  \begin{pmatrix} 0 & 0 & 0 & 0 \\
    0 & 0 & 0 & 1 \\ 0 & 0 & 0 & 0 \\ 0 & 1 & 0 & 0
    \end{pmatrix},\\
    \lambda_{12} \equiv  \begin{pmatrix} 0 & 0 & 0 & 0 \\
    0 & 0 & 0 & -\mathrm{j} \\ 0 & 0 & 0 & 0 \\ 0 & \mathrm{j} & 0 & 0
    \end{pmatrix}  ,  ~~
    \lambda_{13} \equiv & \begin{pmatrix} 0 & 0 & 0 & 0 \\
    0 & 0 & 0 & 0 \\ 0 & 0 & 0 & 1 \\ 0 & 0 & 1 & 0
    \end{pmatrix}, ~~
    \lambda_{14} \equiv  \begin{pmatrix} 0 & 0 & 0 & 0 \\
    0 & 0 & 0 & 0 \\ 0 & 0 & 0 & -\mathrm{j} \\ 0 & 0 & \mathrm{j} & 0
    \end{pmatrix},  \\
    \lambda_{15} \equiv & \left(1/\sqrt{6}\right) \begin{pmatrix} 1 & 0 & 0 & 0 \\
    0 & 1 & 0 & 0 \\ 0 & 0 & 1 & 0 \\ 0 & 0 & 0 & -3
    \end{pmatrix} .
\end{split}
\end{equation}

The {standard model} involves the consideration of two independent
$SU(3)$\index{SU(3)@$SU(3)$} groups, one for {colour} and the other
one for {isospin} and {strangeness}; if {generators} $\lambda_1$ to
$\lambda_8$ apply to one of the $SU(3)$\index{SU(3)@$SU(3)$} groups
we can produce the {generators} of the second group by resorting to
the {basis} elements $\sigma_3$ and $\sigma_{04}.$ The new set of 4
{idempotents} is then given by
\begin{equation}
\begin{split}
    \label{eq:idempotent2}
    f_1 = \frac{(1 + \sigma_3)(1 + \sigma_{04})}{4}, ~~
    f_2 = \frac{(1 + \sigma_3)(1 - \sigma_{04})}{4}, \\
    f_3 = \frac{(1 - \sigma_3)(1 - \sigma_{04})}{4}, ~~
    f_4 = \frac{(1 - \sigma_3)(1 + \sigma_{04})}{4}.
\end{split}
\end{equation}
Again a set of $SU(3)$\index{SU(3)@$SU(3)$} {generators} can be
constructed following a procedure similar to the previous one
\begin{equation}
\label{eq:su3gen2}
\begin{split}
    \alpha_1 &= \sigma_{02} (f_1+f_2)= \frac{\sigma_{02}+\sigma_{023}}{2}, \\
    \alpha_2 &= \sigma_{01} (f_1+f_2)= \frac{\sigma_{01}+\sigma_{013}}{2}, \\
    \alpha_3 &= f_1 - f_2= \frac{\sigma_{04}-\sigma_{034}}{2}, \\
    \alpha_4 &= \sigma_2 (f_2 + f_3)= \frac{\sigma_{2}+\sigma_{024}}{2}, \\
    \alpha_5 &= -\sigma_1 (f_2 + f_3)= \frac{-\sigma_{1}-\sigma_{014}}{2}, \\
    \alpha_6 &= \sigma_4 (f_1 + f_3)= \frac{\sigma_{4}-\sigma_{03}}{2}, \\
    \alpha_7 &= \sigma_{012} (f_1 + f_3)= \frac{\sigma_{012}+\sigma_{1234}}{2}, \\
    \alpha_8 &= \frac{f_1 + f_2 - 2 f_3}{\sqrt{3}}= \frac{2 \sigma_3 + \sigma_{04} +
    \sigma_{034}}{2 \sqrt{3}}.
\end{split}
\end{equation}
This new $SU(3)$\index{SU(3)@$SU(3)$} group is necessarily
independent from the first one because its {matrix} representation
involves {matrices} with all non-zero rows/columns, while the group
generated by $\lambda_1$ to $\lambda_8$ uses {matrices} with zero
fourth row/column. In the following section we will discuss which of
the two groups should be associated with {colour}.

At the end of Sec.\ \ref{dynamics}  we used one particular
idempotent to split the wavefunction into left and right spinors and
here we discuss how the different idempotents are related to the
symmetries discussed above, suggesting a relation between
idempotents and the different elementary particles. We have already
established that each set of 4 {idempotents} is generated by a pair
of commuting {unitary} {basis} elements. Let any two such {basis}
elements be denoted as $h_1$ and $h_2;$ then the product $h_3 = h_1
h_2$ is itself a third commuting {basis} element. For consistence we
choose, as before,
\begin{equation}
    h_1 \equiv \sigma_{023},~~~~h_2\equiv \sigma_{014};
\end{equation}
to get
\begin{equation}
    h_3 \equiv \sigma_{1234},
\end{equation}
which commutes with the other two as can be easily verified. The
result of this exercise is the existence of {triads} of commuting
{unitary} {basis} elements but no tetrads of such elements. We are
led to state that a general {unitary} element is a linear
combination of unity and the three elements of one {triad}
\begin{equation}
\label{eq:unitary}
    h = a_0 + a_1 h_1 + a_2 h_2 + a_3 h_3.
\end{equation}
Since $h$ is {unitary} and the three $h_m$ commute we can write
\begin{equation}
\label{eq:uniteq}
\begin{split}
    h^2 =\,& \left[(a_0)^2 + (a_1)^2 + (a_2)^2 + (a_3)^2\right] + 2 (a_0 a_1 - a_2
    a_3) h_1 + \\
    & + 2 (a_0 a_2 - a_1 a_3) h_2 + 2 (a_0 a_3 - a_1 a_2) h_3
    =1
\end{split}
\end{equation}
The only form this equation can be verified is if the term in square
brackets is unity while all the others are zero. We then get a set
of four simultaneous equations with a total of sixteen solutions, as
follows: 8 solutions with one of the $a_\mu$ equal to $\pm 1$ and
all the others zero, 6 solutions with two of the $a_\mu$ equal to
$-1/2$ and the other two equal to $1/2$ and 2 solutions with all the
$a_\mu$ simultaneously $\pm 1/2.$ The $a_\mu$ coefficients play the
role of {quantum numbers} which determine the particular idempotent
that goes into Eq.\ (\ref{eq:leftright}); these unusual {quantum
numbers} are expressed in terms of the $SU(4)$\index{SU(4)@$SU(4)$}
{generators} $\lambda_3,$ $\lambda_8$ and $\lambda_{15}$ in Table
\ref{t:quantum} in order to highlight the symmetries.
\begin{table}[htb]
{
 \caption{\label{t:quantum} Coefficients for the various
{unitary} elements.}
\begin{center}
\begin{tabular}{ r |r r r|r r r }
  \hline
   $1$ & $\sigma_{023}$  & $\sigma_{014}$   &   $\sigma_{1234}$  & $\lambda_3$
     & $\lambda_8$  & $\lambda_{15}$  \\
   $(a_0)$ & $(a_1)$  &  $(a_2)$  &  $(a_3)$ &  & & \\
  \hline
   $1$ & $0$  & $0$  & $0$  & $0$  & $0$  & $0$
    \\
   $0$ & $1$  & $0$  & $0$  & $0$  & $2/\sqrt{3}$  & $\sqrt{2/3}$ \\
   $0$ & $0$  & $1$  & $0$  & $1$  & $1/\sqrt{3}$  & $-\sqrt{2/3}$  \\
   $0$ & $0$  & $0$  & $1$  & $-1$  & $1/\sqrt{3}$  & $-\sqrt{2/3}$  \\
   $-1$ & $0$  & $0$  & $0$  & $0$  & $0$  & $0$  \\
   $0$ & $-1$  & $0$  & $0$  & $0$  & $-2/\sqrt{3}$  & $-\sqrt{2/3}$   \\
   $0$ & $0$  & $-1$  & $0$  & $-1$  & $-1/\sqrt{3}$  & $\sqrt{2/3}$   \\
   $0$ & $0$  & $0$  & $-1$  & $1$  & $-1/\sqrt{3}$  & $\sqrt{2/3}$   \\
   $-1/2$ & $-1/2$  & $1/2$  & $1/2$  & $0$  & $0$  & $-\sqrt{3/2}$  \\
   $-1/2$ & $1/2$  & $-1/2$  & $1/2$  & $-1$  & $1/\sqrt{3}$  & $1/\sqrt{6}$  \\
   $-1/2$ & $1/2$  & $1/2$  & $-1/2$  & $1$  & $1/\sqrt{3}$  & $1/\sqrt{6}$   \\
   $1/2$ & $-1/2$  & $-1/2$  & $1/2$  & $-1$  & $-1/\sqrt{3}$  & $-1/\sqrt{6}$   \\
   $1/2$ & $-1/2$  & $1/2$  & $-1/2$  & $1$  & $-1/\sqrt{3}$  & $-1/\sqrt{6}$  \\
   $1/2$ & $1/2$  & $-1/2$  & $-1/2$  & $0$  & $0$  & $\sqrt{3/2}$  \\
   $1/2$ & $1/2$  & $1/2$  & $1/2$  & $0$  & $2/\sqrt{3}$  & $-1/\sqrt{6}$   \\
   $-1/2$ & $-1/2$  & $-1/2$  & $-1/2$  & $0$  & $-2/\sqrt{3}$  & $1/\sqrt{6}$   \\
  \hline
\end{tabular}
\end{center}
    }
\end{table}
We don't propose here any direct relationship between the various
idempotents and the known elementary particles, although the fact
that the standard model gauge symmetry group is found as direct
consequence of the monogenic condition which itself generates the
Dirac equation is rather intriguing.

\section{Conclusion and future work}
Monogenic functions applied in the algebra of 5-dimensional
spacetime have been shown to originate laws of fundamental physics
in such diverse areas as relativistic dynamics, quantum mechanics
and electromagnetism, with possible, still unclear, consequences for
cosmology and particle physics. To say that those functions provide
us with a theory of everything is certainly unwarranted at this
stage but it is clear that there is a case for much greater effort
being invested in their study.

There are unanswered questions in the present work. For instance,
how can we avoid an \emph{ad hoc} definition of inertial mass or
what is the true relation between the symmetries generated by
monogenic functions and elementary particles? In spite of its
various loose ends, the formalism is perfectly capable of unifying
relativistic dynamics, quantum mechanics and electromagnetism, which
in itself is no small achievement. Certain developments seem
relatively straightforward but they must be made, even if no knew
predictions are expected. Applying monogenic functions to the
Hydrogen atom should not be difficult because the form of the Dirac
equation we arrived at is perfectly equivalent to the standard one;
one should then find the same solutions but in a GA formalism. In
the same line one could try to solve the equation for a central
gravitational potential, being certain to find quantum states. It is
not clear how important these could be in planetary mechanics or
galaxy dynamics.

Gravitational waves are predicted by the monogenic function
formalism as we pointed out but did not investigate. How important
are they and what chance is there of them being detected by
experiment? We don't know the answer and we don't know what
difficulties lie on the path of those who try to solve the
equations; this is an open area. The sources tensor must be clearly
understood and directly related to geometry; at the moment all
densities, mass, electromagnetic energy, etc. must be inserted in
the equations but one would expect that a perfect theory would
produce such densities out of nothing. In previous papers we
suggested that a recursive, non-linear, equation could be the answer
to the problem but the concept has not yet been formalized and there
are no clear ideas for achieving such goal.

In conclusion, the present work opens the gate of a path that will
possibly lead to an entirely new formulation and understanding of
physics but this path is very likely to have many hurdles to jump
and several dead ends to avoid.

\begin{appendix}
\section{Indexing conventions \label{indices}}
\begin{figure}[htb]
\vspace{11pt}
\centerline{\includegraphics[scale=1.5]{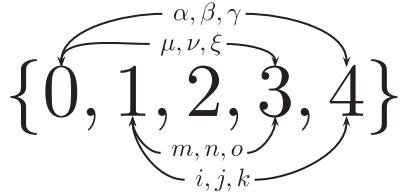}}

\caption{\label{f:indices} Indices in the range $\{0,4\}$ will be
denoted with Greek letters $\alpha, \beta, \gamma.$ Indices in the
range $\{0,3\}$ will also receive Greek letters but chosen from
$\mu, \nu, \xi.$ For indices in the range $\{1,4\}$ we will use
Latin letters $i, j, k$ and finally for indices in the range
$\{1,3\}$ we will use also Latin letters chosen from $m, n, o.$ }
\end{figure}

In this section we establish the indexing conventions used in the
paper. We deal with 5-dimensional space but we are also interested
in two of its 4-dimensional subspaces and one 3-dimensional
subspace; ideally our choice of indices should clearly identify
their ranges in order to avoid the need to specify the latter in
every equation. The diagram in Fig.\ \ref{f:indices} shows the index
naming convention used in this paper;
Einstein's summation convention will be adopted as well as the
compact notation for partial derivatives $\partial_\alpha =
\partial/\partial x^\alpha.$

\section{Time derivative of a 4-dimensional vector}
If there is a refractive index the wave displacement vector can be
written as
\begin{equation}
\label{eq:wavedisp}
    \mathrm{d}x = g_\alpha \mathrm{d}x^\alpha = {n^\beta}_\alpha
    \sigma_\beta \mathrm{d}x^\alpha.
\end{equation}
Because this vector is nilpotent, by virtue of Eq.\
(\ref{eq:nullspecial}), the five coordinates are not independent and
we can divide both members by $\mathrm{d}x^0 = \mathrm{d}t$ defining
the nilpotent vector
\begin{equation}
\label{eq:dotx}
    \dot{x} = g_0 +  g_i \dot{x}^i = {n^\alpha}_0   \sigma_\alpha
    + {n^\beta}_i \sigma_\beta \dot{x}^i.
\end{equation}

Suppose we have a 5D vector $a = \sigma_\alpha a^\alpha$ and we want
to find its time derivative along a path parameterized by $t$, that
is all the $x^i$ are functions of $t$. We can write
\begin{equation}
 \dot{a} = \partial_\beta a^\alpha \dot{x}^\beta \sigma_\alpha;
\end{equation}
where naturally $\dot{x}^0 = 1$. Remembering the definition of
covariant derivative (\ref{eq:covariant}) and Eq.\ (\ref{eq:dotx})
we can modify this equation to
\begin{equation}
\label{eq:timeder}
    \dot{a} = \dot{x}^\beta g_\beta \dprod g^\beta \partial_\beta a^\alpha
    \sigma_\alpha
    = \dot{x} \dprod (\mathrm{D}  a).
\end{equation}
We have expressed vector $a$ in terms of the orthonormed frame in
order to avoid vector derivatives but the result must be independent
of the chosen frame.

This procedure has an obvious dual, which we arrive at by defining
\begin{equation}
    \check{x} = g_\mu \check{x}^\mu + g_4.
\end{equation}
The proper time derivative of vector $a$ is then
\begin{equation}
    \check{a} = \check{x} \dprod (\mathrm{D} a).
\end{equation}

\end{appendix}


\end{document}